\documentclass[conference]{IEEEtran} % 12pt, onecolumn

\pdfoutput=1
\IEEEoverridecommandlockouts
\usepackage{xr-hyper}
\usepackage{acronym}
\usepackage{amsmath}
\usepackage{amssymb}
\usepackage[english]{babel}
\usepackage{caption}
\usepackage{cite}
\usepackage[shortlabels,inline]{enumitem}
\usepackage{etoolbox}
\usepackage{graphicx}
\usepackage[hidelinks]{hyperref}
\usepackage{bookmark}
\usepackage{ifthen}
\usepackage{cleveref}
\usepackage[utf8]{inputenc}
\usepackage{lipsum}
\usepackage{siunitx}
\usepackage[dvipsnames]{xcolor}
\usepackage[T1]{fontenc}
\usepackage[subnum]{cases}
\usepackage{subcaption}
\usepackage{amsthm}
\newtheorem{remark}{Remark}
\usepackage{textcomp}
\usepackage{microtype}
\usepackage{soul}

\usepackage{epstopdf}
\usepackage[toc,symbols]{glossaries}
\usepackage{bm}

\usepackage{algorithm}
\usepackage{algorithmic}
\usepackage{soul}
\usepackage{clipboard}

\newenvironment{change}{\begingroup\color{black}}{\endgroup}
\newcommand{\highlightref}[1]{\expandafter\newcommand\expandafter*\csname bibitem@#1\endcsname{black}}
\makeatother
\highlightref{wang2023music}
\highlightref{wang2021efficient}
\highlightref{wang2021joint}
\highlightref{teng2023variational}
\highlightref{teng2023bayesian}
\highlightref{fascista2022risaided}
\highlightref{sharma2011alpha}
\highlightref{stoica1989music}
\highlightref{stoica1990music}
\highlightref{kay1993fundamentals}
\highlightref{liu2024hierarchical}

\makeatletter
\long\def\@makecaption#1#2{\ifx\@captype\@IEEEtablestring%
    \footnotesize\begin{center}{\normalfont\footnotesize #1}\\
        {\normalfont\footnotesize\scshape #2}\end{center}%
    \@IEEEtablecaptionsepspace
    \else
    \@IEEEfigurecaptionsepspace
    \setbox\@tempboxa\hbox{\normalfont\footnotesize {#1.}~~ #2}%
    \ifdim \wd\@tempboxa >\hsize%
    \setbox\@tempboxa\hbox{\normalfont\footnotesize {#1.}~~ }%
    \parbox[t]{\hsize}{\normalfont\footnotesize \noindent\unhbox\@tempboxa#2}%
    \else
    \hbox to\hsize{\normalfont\footnotesize\hfil\box\@tempboxa\hfil}\fi\fi}
\makeatother

\newacronym{iIRS}{IRS}{intelligent reflecting surface}
\newacronym{iSNR}{SNR}{signal-to-noise ratio}
\newacronym{immWave}{mmWave}{millimeter wave}
\newacronym{iUT}{UT}{user tracking}
\newacronym{iLoS}{LoS}{line-of-sight}
\newacronym{iGLRT}{GLRT}{generalized likelihood ratio test}
\newacronym{iFI}{FI}{Fisher information}
\newacronym{iCSI}{CSI}{channel state information}
\newacronym{iCRB}{CRB}{Cramer-Rao bound}
\newacronym{iMUSIC}{MUSIC}{multiple signal classification}
\newacronym{iML}{ML}{maximum likelihood}

\newacronym{EBA}{EBA}{Extrapolation-Based Algorithm}
\newacronym{FI}{FI}{Fisher information}
\newacronym{SDP}{SDP}{semi-definite programming}
\newacronym{CRB}{CRB}{Cramer-Rao bound}
\newacronym{UT}{UT}{user tracking}
\newacronym{IRS}{IRS}{intelligent reflecting surface}
\newacronym{UE}{UE}{user equipment}
\newacronym{CSI}{CSI}{channel state information}
\newacronym{UAV}{UAV}{Unmanned Aerial Vehicle}
\newacronym{LoS}{LoS}{line-of-sight}
\newacronym{ES}{ES}{Efficient Search Algorithm}
\newacronym{SNR}{SNR}{signal-to-noise ratio}
\newacronym{MSNR}{MSNR}{maximum signal-to-noise ratio}
\newacronym{BS}{BS}{base station}
\newacronym{GA}{GA}{gradient ascent}
\newacronym[\glslongpluralkey={angles of arrival}]{AoA}{AoA}{angle of arrival}
\newacronym[\glslongpluralkey={angles of departure}]{AoD}{AoD}{angle of departure}
\newacronym{mmWave}{mmWave}{millimeter wave}
\newacronym{RF}{RF}{Radio Frequency}
\newacronym{DFT}{DFT}{Discrete Fourier Transform}
\newacronym{MSE}{MSE}{mean square error}
\newacronym{PoV}{PoV}{Point of View}
\newacronym{UPA}{UPA}{uniform planar array}
\newacronym{D}{DT}{data transmission}
\newacronym{CE}{CE}{channel estimation}
\newacronym{IDE}{IDE}{IRS direction estimation}
\newacronym{CS}{CS}{compressed sensing}
\newacronym{SISO}{SISO}{single-input single-output}
\newacronym{NR}{NR}{New Radio}
\newacronym{OFDM}{OFDM}{Orthogonal Frequency-Division Multiplexing}
\newacronym{EKF}{EKF}{Extended Kalman Filter}
\newacronym{B1}{B1}{Baseline 1}
\newacronym{B2}{B2}{baseline 2}
\newacronym{FS}{FS}{full codebook search}
\newacronym{GLRT}{GLRT}{generalized likelihood ratio test}
\newacronym{TB}{TB}{transmission block}
\newacronym{e2e}{e2e}{end-to-end}
\newacronym{pdf}{PDF}{probability density function}
\newacronym{SCA}{SCA}{successive convex approximation}
\newacronym{UC}{UC}{user configuration}
\newacronym{iid}{i.i.d.}{independent and identically distributed}
\newacronym{QoS}{QoS}{quality-of-service}
\newacronym{MUSIC}{MUSIC}{multiple signal classification}
\newacronym{ML}{ML}{maximum likelihood}
\newacronym{MIMO}{MIMO}{multiple-input multiple-output}

\newcommand{\pue}{\ensuremath{\mathbf{p}_\mathrm{UE}}}

\renewcommand{\vec}[1]{\bm{\mathrm{#1}}}
\newcommand{\MIDE}{\mathcal{M}_\mathrm{IDE}}
\newcommand{\TIDE}{T_\mathrm{IDE}}
\newcommand{\MDT}{\mathcal{M}_\mathrm{DT}}
\newcommand{\Mk}{\mathcal{M}_k}
\newcommand{\Mkp}{\mathcal{M}_{k'+1}}
\newcommand{\mideopt}{\mathbf{m}^{\mathrm{IDE}}_{k'+1}}
\newcommand{\omegaM}{\vec{\omega}_{\mathcal{M}}}
\DeclareMathOperator*{\argmin}{arg\,min}
\renewcommand{\baselinestretch}{0.975}

\allowdisplaybreaks

\newclipboard{output_clipboard}
\makeatletter
\let\myorg@bibitem\bibitem
\def\bibitem#1#2\par{%
	\@ifundefined{bibitem@#1}{%
		\myorg@bibitem{#1}#2\par
	}{%
		\begingroup
		\color{\csname bibitem@#1\endcsname}%
		\myorg@bibitem{#1}#2\par
		\endgroup
	}%
}

\makeatother
\newcommand{\copyablelabel}[1]{\label{#1}}
\newcommand{\copyablefigurecaptionpluslabel}[2]{\caption{#1}\label{#2}}

\allowdisplaybreaks

\begin{document}
% \pagenumbering{arabic}
\title{User Tracking and Direction Estimation Codebook Design for IRS-Assisted mmWave Communication}

% \author{Moritz Garkisch$^*$, Sebastian Lotter$^*$, Gui Zhou$^*$, Vahid Jamali$^\dagger$, and Robert Schober$^*$ \\ $^*$\textit{Friedrich-Alexander-Universität Erlangen-Nürnberg}, $^\dagger$\textit{Technische Universität Darmstadt}
\author{Moritz Garkisch, \IEEEmembership{Graduate Student Member, IEEE}, Sebastian Lotter, \IEEEmembership{Member, IEEE}, Gui Zhou, \IEEEmembership{Member, IEEE},\\ Vahid Jamali, \IEEEmembership{Member, IEEE}, Robert Schober, \IEEEmembership{Fellow, IEEE}
% \makeatletter
%\newcommand{\linebreakand}{%
%  \end{@IEEEauthorhalign}
%  \hfill\mbox{}\par
%  \mbox{}\hfill\begin{@IEEEauthorhalign}
%}
% \makeanother
%\author{
  %\IEEEauthorblockN{1\textsuperscript{st} Moritz Garkisch}
  %\IEEEauthorblockA{\textit{Friedrich-Alexander-Universität} \\
    %Erlangen-Nürnberg, Germany \\
    %moritz.garkisch@fau.de}
  %\and
  %\IEEEauthorblockN{2\textsuperscript{nd} Sebastian Lotter}
  %\IEEEauthorblockA{\textit{Friedrich-Alexander-Universität} \\
    %Erlangen-Nürnberg, Germany \\
    %sebastian.g.lotter@fau.de}
  %\and
  %\IEEEauthorblockN{3\textsuperscript{rd} Gui Zhou}
  %\IEEEauthorblockA{\textit{Friedrich-Alexander-Universität} \\
    %Erlangen-Nürnberg, Germany \\
    %gui.zhou@fau.de}
  %\linebreakand % <------------- \and with a line-break
  %\IEEEauthorblockN{4\textsuperscript{th} Vahid Jamali}
  %\IEEEauthorblockA{\textit{Technische Universität Darmstadt} \\
   % Darmstadt, Germany \\
    %vahid.jamali@tu-darmstadt.de}
  %\and
  %\IEEEauthorblockA{\textit{Friedrich-Alexander-Universität} \\
    %Erlangen-Nürnberg, Germany \\
    %robert.schober@fau.de}
\vspace{-7mm}
\thanks{This work was partly supported by the Federal Ministry of Education and Research of Germany under the programme of “Souveran. Digital. Vernetzt.” joint project 6G-RIC (project identification number: PIN 16KISK023) and by the DFG under project number SCHO 831/15-1. Jamali’s work was supported in part by the Deutsche Forschungsgemeinschaft (DFG, German Research Foundation) within the Collaborative Research Center MAKI (SFB 1053, Project-ID 210487104) and in part by the LOEWE initiative (Hesse, Germany) within the emergenCITY center. This work was presented in part at the IEEE ICASSP 2023 (DOI: 10.1109/ICASSP49357.2023.10097027) \cite{garkisch2023codebookbased}.
M. Garkisch, S. Lotter, G. Zhou, and R. Schober are with the Institute for Digital Communication, Friedrich-Alexander-Universität Erlangen-Nürnberg, 91052 Erlangen, Germany (e-mail: \{moritz.garkisch, sebastian.g.lotter, gui.zhou, robert.schober\}@fau.de).
V. Jamali is with the Resilient Communication Systems, Technische Universität Darmstadt, 64283 Darmstadt, Germany (e-mail:vahid.jamali@tu-darmstadt.de).
}
\vspace{-3mm}
}

\maketitle
\thispagestyle{plain}
\pagestyle{plain}

%%%%%%%%%%%%%%%%
%%%%%%%%%%%%%%%%
%%%%%%%%%%%%%%%%
\begin{abstract}
Future communication systems are envisioned to employ \glspl{iIRS} and the \gls{immWave} frequency band to provide reliable high-rate services. For mobile users, the time-varying \gls{iCSI} requires adequate adjustment of the reflection pattern of the \gls{iIRS}. We propose a novel codebook-based \gls{iUT} algorithm for \gls{iIRS}-assisted \gls{immWave} communication, allowing suitable reconfiguration of the \gls{iIRS} unit cell phase shifts, resulting in a high reflection gain. The presented algorithm acquires the direction information of the user based on a peak \gls{iML}-based direction estimation. Using the direction information, the user's trajectory is extrapolated to proactively update the adopted codeword and adjust the \gls{iIRS} phase shift configuration accordingly. Furthermore, we conduct a theoretical analysis of the direction estimation error and utilize the obtained insights to design a codebook specifically optimized for direction estimation.
\Copy{abstract_music}{\begin{change}Our results show that the proposed \gls{iML}-based direction estimation algorithm outperforms a \gls{iMUSIC}-based reference scheme.
The proposed direction estimation codebook improves the direction estimation error for both these schemes as compared to when a reference codebook is used.\end{change}}
\Copy{abstract_kalman}{\begin{change}Also, the accuracy of the proposed \gls{iUT} algorithm is shown to be competitive with Kalman filter-based \gls{iUT}, while the proposed scheme requires fewer {\em a priori} assumptions on the user movement.\end{change}}
Furthermore, the average achieved \gls{iSNR} as well as the average effective rate of the proposed \gls{iUT} algorithm are analyzed. The proposed \gls{iUT} algorithm requires only a low overhead for direction and channel estimation and avoids outdated \gls{iIRS} phase shifts. Furthermore, it is shown to outperform three benchmark schemes based on direct phase shift optimization, optimal codeword selection, and hierarchical codebook search, respectively, via computer simulations.

\end{abstract}
\begin{IEEEkeywords}
Intelligent reflecting surface (IRS), millimeter wave, user tracking, codebook design.
\end{IEEEkeywords}
\glsresetall

%%%%%%%%%%%%%%%%%%%%%%
%%%%%%%%%%%%%%%%%%%%%%
%%%%%%%%%%%%%%%%%%%%%%
\section{Introduction}\copyablelabel{sec:introduction}
The continuous development of new communication systems, such as the sixth-generation (6G) wireless system, aims to provide higher data rates to accommodate the steadily increasing user demands. To meet these requirements, previously unused spectrum, such as the \gls{mmWave} band (30 GHz - 300 GHz), is considered of great interest \cite{samimi2016mmwave}. However, for \gls{mmWave} communication systems, the \gls{LoS} path is usually significantly stronger than the remaining multi-path components \cite{samimi2016mmwave}. Therefore,  obstruction of the \gls{LoS} causes severe performance degradation in these systems. To mitigate this effect, \glspl{IRS} have been introduced. These are passive devices consisting of many small programmable unit cells that reflect electromagnetic waves in a controllable manner \cite{wu2020magazine,wu2021tutorial}. By placing \glspl{IRS} strategically, the coverage of a \gls{BS} can be extended to otherwise obstructed areas, by providing a virtual \gls{LoS} from the \gls{BS} to a user via the \gls{IRS} \cite{wu2020magazine,wu2021tutorial}.

When an \gls{IRS} is utilized to enable communication between a \gls{BS} and a mobile user, the phase shifts of the \gls{IRS} unit cells need to be updated regularly according to the user's current \begin{change}direction\end{change}. This task can be facilitated by \textit{\gls{UT}} techniques that update the \gls{BS}'s estimate of the real-time direction of the user relative to the \gls{IRS}.
In particular, once the direction of the user (or an estimate of it) is available to the \gls{BS}, the \gls{IRS} phase shifts can be configured efficiently via \textit{codebook}-based \gls{IRS} configuration \cite{najafi2020physicsbased}.
Compared to configuring the phase shifts of the \gls{IRS} unit cells individually \cite{abeywickrama2020intelligent,wu2020beamforming}, which requires full \gls{CSI} knowledge, it suffices for codebook-based \gls{IRS} configuration to know the direction of the user in order to select the optimal codeword, i.e., the codeword whose beam pattern overlaps most with the actual direction of the user.
Hence, accurate \gls{UT} in \gls{IRS}-assisted systems enables the efficient configuration of \gls{IRS} unit cells for data transmission.

However, acquiring information about the user's \begin{change}direction\end{change} is itself a difficult problem in \gls{IRS}-assisted systems.
For systems without \glspl{IRS}, direction estimation has been studied extensively and powerful digital signal processing algorithms, such as the \gls{MUSIC} algorithm \cite{schmift1986multiple} and several variants of it (see \cite{ko2018performance} for a recent review of algorithms for direction estimation) have been proposed.
The existing direction estimation schemes for systems with \glspl{IRS} can be divided into \gls{MUSIC}-based direction estimation, \gls{ML}-based direction estimation, and \gls{CRB}-based direction estimation (see \cite{stoica1989music,stoica1990music} for a general introduction and comparison of these approaches).
\Copy{MUSIC_introduction_new_justification}{\begin{change}A \gls{MUSIC} algorithm for \gls{IRS}-assisted multi-user direction estimation in one spatial dimensional has been introduced in \cite{wang2023music}. However, since in \cite{wang2023music} random phase shifts were used to configure the \gls{IRS}, leading to small reflection gains, the feasibility of the scheme proposed in \cite{wang2023music} hinges on the assumptions of large antenna gains and comparatively small path losses at the considered carrier frequency of $2.4\mathrm{GHz}$. Conversely, when applied to \gls{mmWave} systems, which experience higher path loss, and a three-dimensional environment with isotropic antennas, as considered in this paper, the random \gls{IRS} configuration from \cite{wang2023music} leads in general to very low \glspl{SNR} and, consequently, unreliable direction estimation performance.\end{change}}\Copy{introduction_ML_reference}{
\begin{change} \Gls{ML}-based direction estimation has been studied for communication systems with and without \gls{IRS} \cite{wang2021joint,wang2021efficient}.\end{change}\Copy{downlink_wang_efficient_joint}{
\begin{change}In \cite{wang2021efficient}, the authors propose an \gls{ML}-based direction estimation scheme in the downlink using omnidirectional beams for an active multi-antenna array, i.e., a unit modulus constraint, as usually required for \gls{IRS} beamforming, is not considered in \cite{wang2021efficient}. On the other hand, the authors of \cite{wang2021joint} proposed an \gls{ML}-based algorithm for \gls{IRS}-assisted direction estimation in the downlink, which, however, uses random \gls{IRS} configurations resulting in low received \glspl{SNR}.\end{change}}}
% However, those algorithms require the continuous measurement of the received signal at an antenna array. Hence, since standard (passive) \glspl{IRS} cannot perform sensing and processing, these algorithms are not applicable in \gls{IRS}-assisted settings.
An algorithm for estimating the user's direction relative to the \gls{IRS} based on minimizing the \gls{CRB} of the direction estimation error was proposed in \cite{wymeersch2020beyond}.
However, this algorithm requires full \gls{CSI} knowledge, which is difficult to obtain for \gls{IRS}-assisted systems in practice.
Furthermore, the scheme proposed in \cite{wymeersch2020beyond} requires adjusting the phase shifts of all \gls{IRS} elements individually, which is computationally challenging for large \glspl{IRS} and may be infeasible in real-time.
\Copy{new_reference_fascista}{\begin{change}The authors of \cite{fascista2022risaided} proposed a downlink direction estimation beam design for a one-dimensional \gls{IRS} based on the \gls{CRB}. However, this design is limited to a fixed beamwidth, which is equivalent to the 3dB beamwidth of a focused beam and thus depends on the size of the \gls{IRS}.\end{change}}

For user tracking, neural network based prediction approaches can be employed for specific scenarios where movement data is available for training the neural network, see, e.g., \cite{wang2019exploring,lu2023vehicle}. However, the requirement for training data limits the applicability of such models, especially when user movement patterns change over time.
\Copy{new_reference_teng}{\begin{change}In \cite{teng2023bayesian}, a message passing algorithm is employed for tracking of a single user in the downlink with a multiple-\gls{IRS} system, utilizing a Markov transition model assuming random and uncorrelated user movements. The approach presented in \cite{teng2023bayesian} was extended to the multi-user case in \cite{teng2023variational}. However, the methods in \cite{teng2023bayesian} and \cite{teng2023variational} are limited to multiple-\gls{IRS} systems and very specific movement scenarios.
\end{change}}
It was also proposed in the literature to track the user's direction using model-based statistical signal processing tools, such as the Kalman filter \cite{zhang2016tracking}, the extended Kalman filter \cite{va2016beamtracking}, or linear extrapolation \cite{stratidakis2020low}.
However, the accuracy of these methods depends on the availability of models that map the measurements to the user's direction parameters \cite{zhang2016tracking,va2016beamtracking} or the strong assumption of linear user trajectories \cite{stratidakis2020low}.
However, for the general case of mobile users that move nonlinearly, i.e., the user's direction is a nonlinear function of time, a mapping from the measurement, i.e., the received signal at either the \gls{BS} or the user, to the user's direction relative to the \gls{IRS} is not known to date.
\Copy{introduction_weaker_comment_about_model_based_approaches}{\begin{change}Furthermore, model-based user tracking methods typically hinge on {\em a priori} assumptions on the parameters of the considered movement model and these parameters are unknown in general and may even change over time.\end{change}}
Finally, the authors of \cite{huang2022roadside} propose a \gls{UT} algorithm for a roadside setting where several \glspl{IRS} are used to maintain \gls{LoS} connections to cars moving along the road. However, the movement trajectories of the cars in \cite{huang2022roadside} are assumed to be linear and, hence, the proposed scheme does not apply in general, when nonlinear movement of user devices is considered. To the best of the authors' knowledge, a \gls{UT} algorithm for general user movement scenarios which is applicable when {\em full \gls{CSI}}, i.e., knowledge of all multi-path components, is not available has not been proposed for \gls{IRS}-assisted systems, so far.

In this paper, we propose a novel scheme for user direction estimation and tracking in \gls{IRS}-assisted systems to overcome the shortcomings of existing schemes.
On the one hand, the \gls{UT} scheme proposed in this paper is codebook-based and, hence, features efficient \gls{IRS} configuration supporting real-time use cases.
While existing codebooks are either of fixed size \begin{change}\cite{zheng2020irsofdma,fascista2022risaided}\end{change}, i.e., of limited flexibility, or optimized for data transmission, i.e., to achieve the highest possible reflection gains, \cite{najafi2020physicsbased,jamali2020power, ghanem2022optimizationbased,kim2022learningbased}, the {\em direction estimation codebook} proposed in this paper is of variable size and designed specifically for direction estimation.
On the other hand, the proposed tracking scheme does not depend on the assumption of linear user trajectories, but is applicable to general nonlinear user movement.
The proposed algorithm repeatedly estimates the user direction in regular time intervals using a peak \gls{ML}-based approach facilitated by the proposed direction estimation codebook. Then, based on these direction estimates, the user's trajectory is extrapolated in the subsequent time blocks and the corresponding optimal codewords for data transmission are selected from a second codebook, the {\em data transmission codebook}.
In order to design the direction estimation codebook, first a theoretical analysis of the expected estimation error using a first-order approximation of the \gls{IRS} response function is performed.
Then, based on the error analysis, a gradient descent algorithm for finding a locally optimal beam shape is derived.

The main contributions can be summarized as follows:
\begin{itemize}
    \item We propose a novel beam pattern design specifically optimized for minimizing the user direction estimation error in the considered \gls{IRS}-assisted system. The direction estimation codebook derived from the proposed beam pattern is practically realizable since it implicitly accounts for beam shape limitations imposed by the codebook-based \gls{IRS} configuration.
    \item We introduce a codebook-based \gls{UT} algorithm that first estimates the direction based on the proposed direction estimation codebook and then extrapolates the current user direction. The extrapolated direction is used to proactively determine when the \gls{IRS} codeword for data transmission has to be changed and to optimally select the new codeword.
    \item We provide a comprehensive analysis of the overall performance and overhead that the proposed communication scheme achieves.
\end{itemize}
% new order (suggested by GZ)
\Copy{introduction_music_comparison}{\begin{change}The simulation results reveal that the proposed \gls{ML}-based direction estimation method outperforms a \gls{MUSIC} algorithm baseline for all considered \gls{IRS} phase shift configurations.\end{change}}
Also, the direction estimation error is significantly lower when employing the proposed direction estimation codebook, compared to a reference codebook from the literature optimized for data transmission. 
The results confirm that the proposed \gls{UT} algorithm can maintain the \gls{LoS} connection to a mobile user, even for nonlinear movement patterns.
\Copy{introduction_kalman_comparison}{\begin{change}The tracking accuracy of the proposed trajectory-based algorithm is equivalent to a Kalman filter baseline, while requiring fewer assumptions about the user movement.\end{change}}
Overall, the proposed \gls{UT} algorithm achieves a significantly higher effective rate compared to three baseline schemes, which are based on the perfect \gls{CSI} and codebook-based \gls{IRS} configuration for data transmission via hierarchical search and optimal codeword selection, respectively.

In the conference version of this paper \cite{garkisch2023codebookbased}, we introduced a preliminary version of the \gls{UT} algorithm presented in this paper. However, the codebook used for direction estimation in \cite{garkisch2023codebookbased} was not optimized and a comprehensive performance analysis of the proposed scheme was not provided in \cite{garkisch2023codebookbased}. Hence, this paper significantly extends \cite{garkisch2023codebookbased} towards a mature \gls{UT} scheme for \gls{IRS}-assisted mmWave systems.

The remainder of this paper is organized as follows. In Section \ref{sec:system_model}, we introduce the considered system model. The \gls{UT} algorithm and the codebook design are presented in Sections \ref{sec:tracking_scheme} and \ref{sec:codebook_generation}, respectively. Finally, Sections V and VI report our numerical results and conclusions, respectively.  

\textit{Notations}: Lower case and upper case bold letters denote vectors and matrices, respectively. The transpose and conjugate transpose of matrix $\mathbf{A}$ are denoted by $\mathbf{A}^{\mathrm{T}}$ and $\mathbf{A}^{\mathrm{H}}$, respectively. The $i$-th element of vector $\mathbf{a}$ is denoted by $[\mathbf{a}]_{i}$, and the element in the $i$-th row and $j$-th column of matrix $\mathbf{A}$ is denoted by $[\mathbf{A}]_{i,j}$. The distribution of a circularly symmetric complex Gaussian random vector with mean vector $\mathbf{x}$ and covariance matrix $\mathbf{A}$ is represented by $\mathcal{CN}(\mathbf{x},\mathbf{A})$ and the imaginary unit is denoted by $\mathrm{j} = \sqrt{-1}$. The all-zero vector of length $n$, the all-one vector of length $n$, and the identity matrix of size $n\times n$ are denoted as $\mathbf{0}_n$, $\mathbf{1}_n$, and $\mathbf{I}_n$, respectively. The complex conjugate, absolute, and expected values of a scalar $x$ are denoted by $x^*$, $|x |$, and $\mathcal{E}\{x\}$, respectively. The main diagonal vector and the rank of a matrix $\mathbf{X}$ are denoted as $\mathrm{Diag}(\mathbf{X})$ and $\mathrm{Rank}(\mathbf{X})$, respectively. The Hadamard (element-wise) and Kronecker products are denoted by $\odot$ and $\otimes$, respectively. The cardinality of set $\mathcal{M}$ is denoted by $|\mathcal{M}|$. $\mathbb{N}$, $\mathbb{R}$, $\mathbb{R}_+$, and $\mathbb{C}$ denote the sets of natural, real, positive real, and complex numbers, respectively. For any real number $x$, $\lfloor x \rfloor$ denotes the floor of $x$. Finally, the big-O notation is denoted by $\mathcal{O}(\cdot)$.

%%%%%%%%%%%%%%%%%%%%%%
%%%%%%%%%%%%%%%%%%%%%%
%%%%%%%%%%%%%%%%%%%%%%
\section{System Model}\copyablelabel{sec:system_model}
\begin{figure}
	\centering
    \includegraphics[clip, trim=3cm 6.3cm 0cm 5.7cm,width=0.45\textwidth]{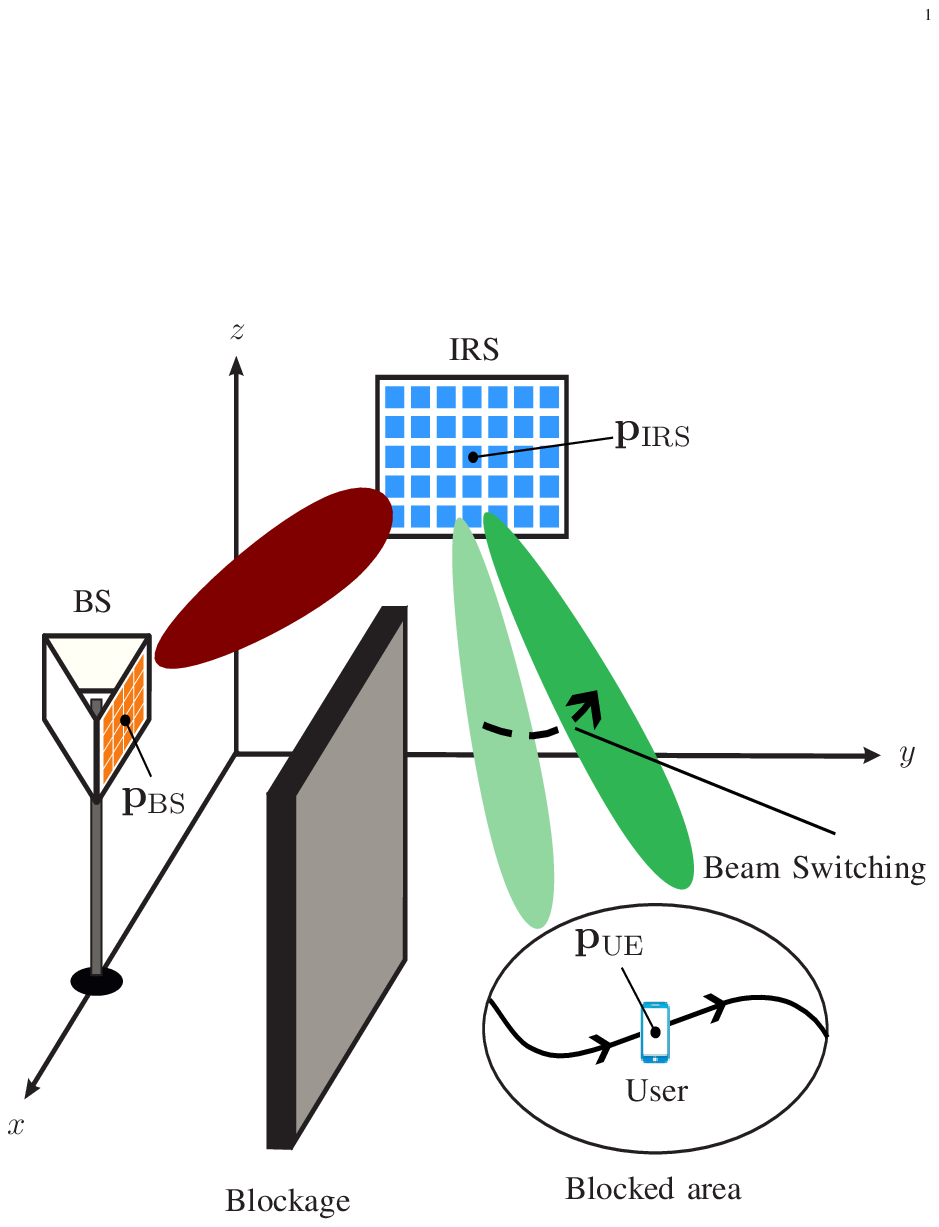}
    \caption{The considered system consists of a \gls{BS}, an \gls{IRS}, and a user that moves within an obstructed area. The direct link between the \gls{BS} and the user is blocked. } \label{fig:system_model}
    \vspace{-5mm}
\end{figure}
% \todo[inline,color=red]{In Fig.~\ref{fig:system_model}, the color of the BS beam is darker now and the path is nonlinear. Gui want s to change the BS array to make it look more differnt than the IRS.}

We consider the \gls{IRS}-assisted \gls{mmWave} communication system illustrated in Fig.~\ref{fig:system_model}.
In this \begin{change}downlink\end{change}\footnote{\Copy{tell_alternative_uplink}{\Copy{footnote_alternative_uplink_part_1}{\begin{change}
    Alternatively, user tracking could be performed in the uplink. For a reciprocal channel, the performance would then be equivalent to that of the considered downlink scheme. \Copy{system_model_ul_dl}{In particular, the beamformers at the \gls{BS} and the user and the \gls{IRS} reflection gain impact the end-to-end channel gain and, hence, the direction estimation accuracy, equally in uplink and downlink.} The data transmission can be performed in uplink or downlink, independent of the user tracking. However, for ease of presentation and without loss of generality, we exclusively consider downlink user tracking and downlink data transmission in this paper.
\end{change}}}} system, a \gls{BS} transmits data to a mobile user who is located in an area to which no \gls{LoS} path from the \gls{BS} is available throughout the considered time frame.
The \gls{BS} and the user device are equipped with \glspl{UPA} comprising $Q_\mathrm{BS} = Q_{\mathrm{BS},1} \times Q_{\mathrm{BS},2}$ and $Q_\mathrm{UE} = Q_{\mathrm{UE},1} \times Q_{\mathrm{UE},2}$ elements with spacing $d_\mathrm{BS}$ and $d_\mathrm{UE}$, respectively, and both employ analog beamforming.
We equip the considered communication environment with a Cartesian coordinate system, cf.~Fig.~\ref{fig:system_model}, such that the \gls{BS} \gls{UPA} lies in the $x$-$z$ plane and the \gls{BS} and user \glspl{UPA} are centered at Cartesian coordinates $\mathbf{p}_\mathrm{BS} \in \mathbb{R}^3$ and $\pue(t) \in \mathbb{R}^3$, respectively.
Both $\pue(t)$ and the orientation of the user \gls{UPA} change as functions of time $t$ and are not known to the \gls{BS}.

The situation illustrated in Fig.~\ref{fig:system_model} in which no \gls{LoS} between \gls{BS} and user is available commonly arises in \gls{mmWave} systems due to the pronounced attenuation of high-frequency signals.
To mitigate this lack of a direct signal propagation path, a passive \gls{IRS} to which a \gls{LoS} link from the \gls{BS} is available is employed to assist the communication.
The \gls{IRS} comprises $Q_\mathrm{IRS} = Q_{\mathrm{IRS},1} \times Q_{\mathrm{IRS},2}$ unit cells of sizes $A_\mathrm{UC} = d_\mathrm{IRS}^2$ and is centered at $\mathbf{p}_\mathrm{IRS} = [x_\mathrm{IRS},y_\mathrm{IRS},z_\mathrm{IRS}]^{\mathrm{T}} \in \mathbb{R}^3$.
For simplicity of presentation and without loss of generality, in this paper, we assume that the \gls{IRS} is square, i.e., $Q_{\mathrm{IRS},1} = Q_{\mathrm{IRS},2} = Q$, and lies in the $y$-$z$ plane, i.e., orthogonal to the \gls{BS} \gls{UPA}\footnote{The latter assumption is not critical for the theory developed in this paper. It is introduced mainly for notational convenience and can be removed by a slight generalization of the notation.}.

In principle, the direction of the user relative to the \gls{IRS} at time $t$ can be described in terms of the azimuth and elevation angles of $\pue(t)$.
However, for the problems of estimating and tracking the direction of the user studied in this paper, it is beneficial instead to consider the direction of $\pue(t)$ relative to the line $\mathbf{p}_\mathrm{IRS} + a [1,0,0]^{\mathrm{T}}$, $a \in \mathbb{R}_+$, i.e., the normal direction of the \gls{IRS}.
To this end, we introduce the notation $[\theta,\phi]^\mathrm{T}$, where $\theta$ and $\phi$ for any $[x,y,z] \in \mathbb{R}_+ \times \mathbb{R}^2$ are defined as $\theta = \arctan\left(\frac{y-y_\mathrm{IRS}}{x}\right)$ and $\phi = \arctan\left(\frac{z-z_\mathrm{IRS}}{x}\right)$. Furthermore, whenever it is necessary to consider the azimuth and elevation angles in the following, we use the classical definition, where the elevation angle $\epsilon$ is the angle between the direction vector and the normal vector of the respective array, and the azimuth angle $\alpha$ is the angle between the horizontal and vertical axis of the direction vector projection onto the array plane.

\subsection{Time Block Structure}\copyablelabel{sec:system_model:time_block_structure}
\vspace{-1mm}
\Gls{BS}, \gls{IRS}, and user communicate in the considered system in {\em \glspl{TB}} of fixed length $T$.
Hereby, the \glspl{TB} are enumerated by $k \in \mathbb{N}_0$ and the starting time of each \gls{TB} $k\in\mathbb{N}_0$ is denoted as $t_k = kT$.
The \gls{TB} structure proposed in this paper, which is identical for all \glspl{TB}, is shown in Fig.~\ref{image:block_structure}.
In the first sub-block of length $T_\mathrm{UC}$ of each \gls{TB} $k$, the so-called {\em \gls{UC} sub-block}, the user's beamformer is updated.
Regularly updating the user's beamformer is necessary to account for the time-varying user \begin{change}direction\end{change} and user \gls{UPA} orientation.
However, since the number of antennas of the user device is typically small in practical systems, the user \gls{UPA}'s beams are rather wide and, consequently, a single update of the user's beamformer per \gls{TB} is considered sufficient.
The second sub-block of length $T_\mathrm{IDE}$ of each \gls{TB}, called {\em \gls{IDE} sub-block}, is used for estimating the direction of the user relative to the \gls{IRS}.
In the \gls{IDE} sub-block, the \gls{BS} repeatedly sends a pilot sequence, while the \gls{IRS} cycles through different \gls{IRS} beam patterns, i.e., codewords.
Based on the received signals for the different \gls{IRS} configurations, the user performs the direction estimation (and feeds the estimated direction back to the \gls{BS}). The remaining part of each \gls{TB} consist of $\eta \in \mathbb{N}$ pairs of {\em \gls{CE} sub-blocks} and {\em \gls{D} sub-blocks}, where each \gls{CE} and \gls{D} sub-block is of length $T_\mathrm{CE}$ and $T_\mathrm{DT}$, respectively.
Since the end-to-end communication channel between \gls{BS} and user is affected by small-scale fading, the end-to-end \gls{CSI} is estimated in the \gls{CE} sub-blocks on the timescale of the channel coherence time $T_\mathrm{coh}$ to facilitate data detection in the \gls{D} blocks. We note that since in the considered system an \gls{LoS} is always available from the \gls{BS} to the \gls{IRS}, the \gls{BS}'s beam can be statically aligned with the \gls{LoS} direction towards the \gls{IRS}. Since no reconfiguration is needed during the system's operation, no time resources need to be allocated in the \gls{TB} structure for configuring the \gls{BS}.

Considering the proposed \gls{TB} structure, two different time scales are employed for \gls{CE} on the one hand and \gls{UC} and \gls{IDE} on the other hand.
Specifically, similar to \cite{alexandropoulous2022nearfield}, \gls{CE} is carried out frequently, while \gls{UC} and \gls{IDE} are performed comparatively infrequently\footnote{Since the end-to-end channel is a scalar value, the overhead and complexity required for \gls{CE} are low, see Section \ref{sec:overhead} for details.}. The duration of one \gls{TB} is $T = T_\mathrm{UC} + T_\mathrm{IDE} + \eta (T_\mathrm{CE} + T_\mathrm{DT})$. To ensure that the end-to-end channel is quasi static during \gls{UC}, \gls{CE}, and \gls{D}, the \gls{IRS} configuration can only be changed directly {\em before} each \gls{UC} sub-block at times $t=t_k$ and directly {\em before} each \gls{CE} sub-block at times $t_{k,\kappa}= t_k + T_\mathrm{UC} + T_\mathrm{IDE} + \kappa (T_\mathrm{CE}+T_\mathrm{DT})$, $\kappa = 0,...,\eta-1$, but not {\em within} any \gls{UC}, \gls{CE}, or \gls{D} sub-block.
Furthermore, the \gls{IRS} phase shift configurations are drawn from the direction estimation codebook $\MIDE$ for the \gls{IDE} sub-blocks and from the data transmission codebook $\MDT$ for all other sub-blocks. For simplicity, we assume that the \gls{IRS} can be reconfigured instantly without any delay 
\Copy{BS_IRS_control_link}{\begin{change} and that a low-rate control link is used by the \gls{BS} to configure the \gls{IRS} according to the \gls{UT} and data transmission scheme detailed in the following sections.\end{change}}

\subsection{User Tracking}\label{sec:system_model:user_tracking}
Since the \gls{IRS} typically comprises many unit cells, it can create narrow reflection beams, e.g., as compared to the user's \gls{UPA} beams.
Consequently, the received power at the user is very low when the \gls{IRS} beam is not aligned with the user direction.
Hence, the \gls{IRS} beam pattern needs to be updated even when the direction of the user changes only slightly and, to this end, it is beneficial if the \gls{BS} keeps track of the user's current \begin{change}direction\end{change}.

In the user tracking scheme proposed in this paper, the \gls{IRS} beam is adapted to changes in the user's direction also {\em within} one \gls{TB} in order to avoid huge drops in the received signal power due to beam misalignment between two consecutive \gls{IDE} blocks.
To this end, the \gls{BS} extrapolates the user's current direction at each $t_{k,\kappa}$, $\kappa=1,\ldots,\eta-1$, based on the direction estimates obtained in the \gls{IDE} sub-blocks of previous \glspl{TB} $k' < k$.
In this way, the \gls{IRS} phase shifts are reconfigured frequently and changes in the user's \begin{change}direction\end{change} are accounted for within the same \gls{TB}.

\Copy{fig:copy_block_structure}{
\begin{figure}
    \centering
    \includegraphics[clip, trim=5.2cm 17.5cm 0.5cm 4.5cm,width=0.45\textwidth]{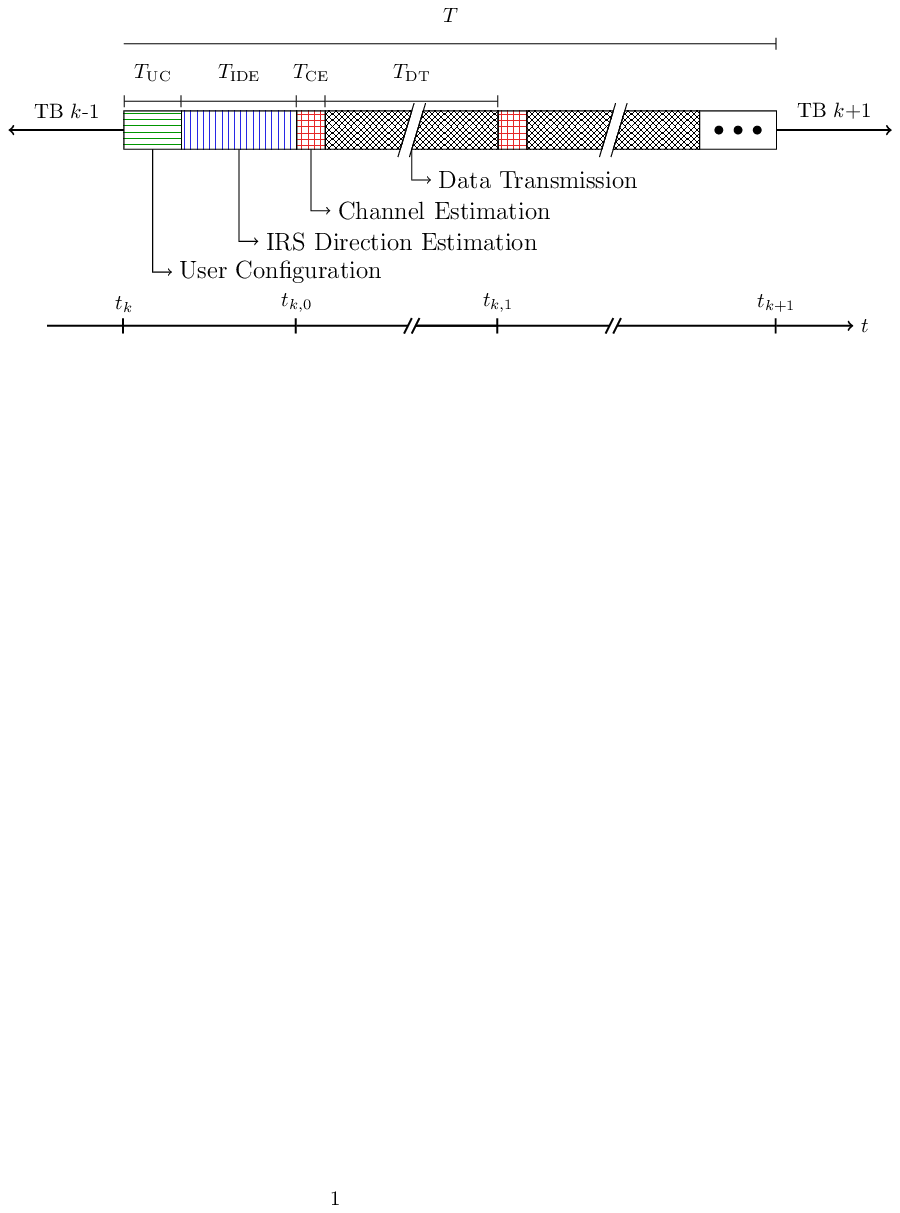}
    \vspace{-3mm}
    \copyablefigurecaptionpluslabel{Transmission block $k$ in the proposed block structure.}{image:block_structure}
    \vspace{-3mm}
\end{figure}}

%%%%%%%%%%%%%%%%%%%%%%%%%%%%%%%%%%%%%%%%%
\vspace{-1mm}
\subsection{Signal Model}\label{sec:signal_model}
The downlink signal received at the user from the \gls{BS} at any time $t$ is given by \cite{najafi2020physicsbased}
\begin{equation}\copyablelabel{eq:tx_model_H}
    r(\mathbf{m},t) = \mathbf{f}_\mathrm{UE}^\mathrm{H}[k] \Big( \mathbf{H}_\mathrm{r} \mathbf{\Omega}(\mathbf{m}) \mathbf{H}_\mathrm{t} \mathbf{f}_\mathrm{BS} s(t) + \mathbf{n}(t) \Big ) ,
\end{equation}
where $\mathbf{f}_\mathrm{UE}[k] \in \mathbb{C}^{Q_\mathrm{UE} }$, $\mathbf{f}_\mathrm{BS}\in \mathbb{C}^{Q_\mathrm{BS} }$, $s(t)$, and $\mathbf{n}(t) \sim \mathcal{CN}(\mathbf{0}_{Q_\mathrm{UE}},\sigma^2 \mathbf{I}_{Q_\mathrm{UE}}) $ denote the beamforming vector at the user in \gls{TB} $k = \lfloor t/T \rfloor$, the beamforming vector at the \gls{BS}, the \gls{BS} transmit signal, and the additive white Gaussian noise (AWGN) at the user, respectively. Furthermore, $\mathbf{H}_\mathrm{r} \in \mathbb{C}^{Q_\mathrm{UE} \times Q_\mathrm{IRS}}$ and $\mathbf{H}_\mathrm{t} \in \mathbb{C}^{Q_\mathrm{IRS} \times Q_\mathrm{BS}}$ represent the \gls{IRS}-to-user and the \gls{BS}-to-\gls{IRS} channels, respectively. Moreover, we define the end-to-end channel as $h_\mathrm{e2e} = \mathbf{f}_\mathrm{UE}^\mathrm{H}[k] \mathbf{H}_\mathrm{r} \mathbf{\Omega}(\mathbf{m}) \mathbf{H}_\mathrm{t} \mathbf{f}_\mathrm{BS}$, where diagonal matrix $\mathbf{\Omega}(\mathbf{m})$ contains the reflection coefficients of all \gls{IRS} unit cells for codeword $\mathbf{m}$ and $\vec{m}$ is drawn from either $\MDT$ or $\MIDE$. For notational convenience, we further define $\boldsymbol{\omega}(\mathbf{m}) = \mathrm{Diag}(\mathbf{\Omega}(\mathbf{m}))$.

The beamformer at the \gls{BS}, $\mathbf{f}_\mathrm{BS}$, is set to align with the (static) \gls{LoS} direction from the \gls{BS} to the \gls{IRS}.
For the beamformer at the user, $\mathbf{f}_\mathrm{UE}[k]$, we assume that it is obtained from a DFT codebook having a size equal to the number of user antennas $Q_{\mathrm{UE}}$\footnote{For an overview of this and alternative combining approaches see \cite{mabrouki2022codebook}.}. Specifically, in each \gls{UC} block, the \gls{BS} repeatedly sends a pilot sequence while the user cycles through all candidate codewords from the DFT codebook.
During the testing of all candidate codewords for \gls{UC}, the \gls{IRS} unit cell phase shifts are fixed to one single codeword.
Finally, $\mathbf{f}_\mathrm{UE}[k]$ is selected as the codeword that resulted in the highest received power.
For a pilot sequence of $N_\mathrm{UC}$ symbols, each of length $T_\mathrm{S}$, the duration of the \gls{UC} sub-block is $T_\mathrm{UC} = Q_\mathrm{UE} N_\mathrm{UC} T_\mathrm{S}$.

%%%%%%%%%%%%%%%%%%%%%%%%%%%%%%%%%%%%%%%%%
\subsection{Channel Model}\copyablelabel{sec:channel_model}
The channels considered in this paper comprise only a few multi-path components, due to the high path loss at \gls{mmWave} frequencies.
Furthermore, since the \gls{BS}-to-\gls{IRS} and \gls{IRS}-to-user channels are assumed to be unobstructed, the respective \gls{LoS} paths are dominant.
To account for additional multi-path components, we assume that there are $L_\mathrm{t}$ and $L_\mathrm{r}$ scatterers at fixed positions in the \gls{BS}-to-\gls{IRS} and \gls{IRS}-to-user channels, respectively.
Consequently, the channels can be characterized according to a geometric channel model, where the channel matrices follow a Rician distribution, see also \cite{li2014channelmodel}.

The \gls{BS}-to-\gls{IRS} and \gls{IRS}-to-user channel matrices are constructed as $\mathbf{H}_\mathrm{t} = \mathbf{A}_\mathrm{t} \mathbf{\Sigma}_\mathrm{t} \mathbf{A}_\mathrm{BS}^\mathrm{H}$ and $\mathbf{H}_\mathrm{r} = \mathbf{A}_\mathrm{UE} \mathbf{\Sigma}_\mathrm{r} \mathbf{A}_\mathrm{r}^\mathrm{H}$, respectively \cite{najafi2020physicsbased}. Here, steering matrices $\mathbf{A}_i = [\mathbf{a}(\boldsymbol\psi_{i,1}),...,\mathbf{a}(\boldsymbol\psi_{i,L_i+1})]$, $i\in\{ \mathrm{UE}, \mathrm{t}, \mathrm{r}, \mathrm{BS} \}$, contain the {\em steering vectors} $\mathbf{a}(\boldsymbol\psi_{i,l})$ for the angle of arrival ($i = t, \mathrm{UE}$) or the angle of departure ($i = \mathrm{BS}, r$) $\boldsymbol\psi_{i,l}, l\in \{ 1 , ..., L_i+1\}$ of propagation path $(i,l)$, and diagonal matrices $\mathbf{\Sigma}_i, i \in \{ \mathrm{t}, \mathrm{r} \}$, contain the path gains associated with channel $i$.

For any device $I\in\{\mathrm{BS},\mathrm{IRS},\mathrm{UE}\}$, every direction relative to $I$ is defined as $\vec{\psi}_I = [\alpha_I, \epsilon_I]$, where $\alpha_I$  and $\epsilon_I$ denote the corresponding azimuth and elevation angles, respectively. The steering vector for $\vec{\psi}_I$ is then given by \cite{jamali2020power}\begin{equation}\copyablelabel{eq:steering_vector}
    \mathbf{a}(\vec{\psi}_I) = \mathbf{a}_1(\vec{\psi}_I) \otimes \mathbf{a}_2(\vec{\psi}_I), 
\end{equation}
where $[\mathbf{a}_\mu(\vec{\psi}_I) ]_q = \mathrm{e}^{\mathrm{j} \frac{2\pi d_I}{\lambda} A_\mu(\vec{\psi}_I) \times \left( - \frac{Q_{I,\mu}-1}{2} + (q-1) \right)}$, for $q\in\{1,...,Q_{I,\mu}\}$, $\mu\in\{1,2\}$. Here, $\lambda$ denotes the wavelength and $A_1(\vec{\psi}_I)$ $=$ $\sin(\epsilon_I) \cos (\alpha_I)$ and $A_2(\vec{\psi}_I)$ $=$ $\sin (\epsilon_I) \sin (\alpha_I)$ define the phase differences between adjacent array elements in horizontal and vertical direction, respectively. For convenience, and in slight abuse of notation, we use the notations $\mathbf{a}(\vec{\psi}_{\mathrm{IRS}})$ and $\mathbf{a}(\vec{\psi})$ interchangeably, where $\vec{\psi} = [\theta, \phi]$ is given in the $\theta$, $\phi$ coordinates defined above and $[\alpha_{\mathrm{IRS}}, \epsilon_{\mathrm{IRS}}]$ is mapped to $[\theta, \phi]$ as $\alpha_\mathrm{IRS} = \arctan\left(\sqrt{\tan^2(\theta) + \tan^2(\phi)}\right)$ and $\epsilon_\mathrm{IRS} = \arctan\left(\frac{\tan(\theta)}{\tan(\phi)}\right) + \frac{\pi}{2} \left( 1-\mathrm{sign}\left(\tan(\phi)\right) \right)$.
The attenuation of the electromagnetic waves along any propagation path is modeled here as free space path loss, i.e.,
\begin{equation}
    [\mathbf{\Sigma}_i]_{l,l} = \left( \upsilon_{i,l} \frac{\lambda}{4 \pi \delta_{i,l}} \right)^2, \; i\in\{\mathrm{t}, \mathrm{r}\} ,
\end{equation}
where $\upsilon_{i,l}$ and $\delta_{i,l}, l\in\{1,..., L_i+1\}$, denote the reflection coefficient and the distance that the electromagnetic waves travel along path $(i,l)$, respectively. 
Assigning, without loss of generality, index $1$ to the \gls{LoS} directions, e.g., $\boldsymbol\psi_{\mathrm{BS},1}$ corresponds to the \gls{LoS} direction from the \gls{BS} to the \gls{IRS}, the ratio of the gain of the \gls{LoS} path to the cumulative gain of all other paths is denoted by 
\begin{equation}\copyablelabel{eq:rice_factor}
    K_i = \frac{[\mathbf{\Sigma}_i]_{1,1}}{\sum_{l=2}^{L_i+1} [\mathbf{\Sigma}_i]_{l,l} }, \;\; i\in\{\mathrm{t}, \mathrm{r}\} .
\end{equation}
\vspace{-5mm}
\begin{change}\begin{remark}\label{remark:1}
\Copy{remark_scatters}{The proposed direction estimation scheme presented in Section~\ref{sec:tracking_scheme} is based on the assumption that only the \gls{LoS} path is present. However, in order to assess the robustness of the proposed algorithm, we consider scatterers, which are likely present in real-world environments, for the performance evaluation in Section~\ref{sec:performance}.}
\end{remark}\end{change}
\vspace{-3mm}
%%%%%%%%%%%%%%%%%%%%%%%%%%%%%%%%%%%%%%%%%
\subsection{Codebook Model}\label{sec:codebook_model}
We recall from above that $\MIDE$ denotes the codebook used to configure the phase shifts of the \gls{IRS} unit cells in the \gls{IDE} sub-blocks and assume, similar to \cite{jamali2020power,you2020fast}, that the phase shift vectors for the horizontal and vertical directions are adjusted separately. Even though the full flexibility of the \gls{IRS} is not available under this assumption, it is critical in order to reduce the computational complexity of the beam design. Specifically, only $2 Q$ phase shifts need to be optimized as compared to $Q^2$ phase shifts when optimizing the phase shift of each unit cell individually.
Consequently, the phase shift vector corresponding to codeword $\vec{m} \in \MIDE$, $\vec{\omega}(\vec{m})$, factorizes into a horizontal and a vertical component as follows
\begin{equation}\copyablelabel{eq:split_omega}
 \boldsymbol{\omega}(\mathbf{m}) = \boldsymbol{\omega}_1(m_1) \otimes \boldsymbol{\omega}_2(m_2),
\end{equation}
where $\boldsymbol{\omega}_1(m_1), \boldsymbol{\omega}_2(m_2) \in \mathbb{C}^{Q}$ denote the \gls{IRS} phase shift vectors in horizontal and vertical direction, respectively, and we can write $\mathbf{m}=[m_1, m_2]^T$, where $m_1, m_2 \in \{ 0,...,M-1 \}$ and $M$ denotes the codebook size in both horizontal and vertical direction.
According to this codebook construction, $\MIDE$ comprises $|\MIDE|=M^2$ codewords.

To further simplify the codebook design and facilitate real-time applications, in this paper, we employ the same beam shape $\boldsymbol{\omega}_\mathcal{M} \in \mathbb{C}^Q$ for each codeword, such that the presented codebook design problem, cf.~Section~\ref{sec:codebook_generation}, has to be solved only once per codebook and not for each codeword individually.
Consequently, we have
\begin{equation}\copyablelabel{eq:codeword_linear_shift}
\begin{split}
    & \boldsymbol{\omega}_\mu(m_{\mu}) = \omegaM \odot \vec{\omega}_{m_{\mu}} \\
    & \;\; = \boldsymbol{\omega}_\mathcal{M} \odot \left [ \mathrm{e}^{\mathrm{j} \left( \frac{4\pi d_\mathrm{IRS} m_\mu}{\lambda M} - \pi \right) 1},...,\mathrm{e}^{ \mathrm{j} \left( \frac{4\pi d_\mathrm{IRS} m_\mu}{\lambda M} - \pi \right) Q} \right]^\mathrm{T},
\end{split}
\end{equation}
i.e., $\vec{\omega}_{\mathcal{M}}$ is rotated towards direction $\vec{\psi}_{\vec{m}}$, that is defined as the solution to $A_\mu(\boldsymbol\psi_{\vec{m}}) = \frac{2 m_\mu}{M}-\frac{\lambda}{2 d_\mathrm{IRS}}, \mu\in\{ 1,2 \}$, where $\vec{\psi}_{\vec{m}}$ is defined in terms of the $\theta$, $\phi$ coordinates and $A_\mu(\boldsymbol\psi_{\vec{m}})$ has been defined above.

%%%%%%%%%%%%%%%%%%%%%%%%%%%%%%%%%%%%%%%%
%%%%%%%%%%%%%%%%%%%%%%%%%%%%%%%%%%%%%%%%
%%%%%%%%%%%%%%%%%%%%%%%%%%%%%%%%%%%%%%%%
\section{Codebook-Based Direction Estimation and User Tracking}\copyablelabel{sec:tracking_scheme}
In this section, we propose a novel scheme for estimating and tracking the user's direction relative to the \gls{IRS}.
To this end, we first specialize the received signal model in \eqref{eq:tx_model_H} for the transmission in the \gls{IDE} sub-blocks, cf.~Section~\ref{sec:system_model:time_block_structure}.
Then, we investigate how the received signal in the \gls{IDE} sub-blocks can be utilized to estimate the user's current direction and how, based on the direction information from multiple consecutive \gls{IDE} sub-blocks, the user's direction can be extrapolated at times $t_k$ and $t_{k,\kappa}$, cf.~Sections~\ref{sec:system_model:time_block_structure} and \ref{sec:system_model:user_tracking}.
The framework proposed in this section is applicable for any direction estimation codebook $\MIDE$ and a specific design of $\MIDE$ is introduced later in Section~\ref{sec:codebook_generation}.

For further reference, we define the direction in which the reflection gain of the \gls{IRS}, configured with codeword $\mathbf{m}$, for an impinging wave from the \gls{BS} (i.e., from direction $\boldsymbol\psi_{\mathrm{t},1}$), is maximal as
\begin{equation}\copyablelabel{eq:main_lobe_direction}
    \boldsymbol\psi_\mathrm{IRS}(\mathbf{m}) = \begin{bmatrix}
        \theta_\mathrm{IRS}(\vec{m})\\
        \phi_\mathrm{IRS}(\vec{m})
    \end{bmatrix} = \mathrm{arg} \max_{\boldsymbol\psi} | \mathbf{a}^\mathrm{H}(\boldsymbol\psi) \boldsymbol{\Omega}(\mathbf{m}) \mathbf{a}(\boldsymbol\psi_{\mathrm{t},1}) |.
\end{equation}

%%%%%%%%%%%%%%%%%%%%%%%%%%%%%%%%%
\subsection{Measurements for Direction Estimation}\label{sec:measurements_for_direction_estimation}
\Copy{justify_LoS_approximation}{Due to the propagation properties of electromagnetic waves at \gls{mmWave} frequencies, it is often possible to approximate the received signal \eqref{eq:tx_model_H} by only the \gls{LoS} path contribution, while ignoring the remaining multi-path components \cite{samimi2016mmwave}.
We apply this approximation for the theoretical analysis in the following, yet, as pointed out in Remark~\ref{remark:1}, all multi-path components are considered in the performance evaluation of the proposed scheme in Section~\ref{sec:performance}.}
Considering only the \gls{LoS} contribution, \eqref{eq:tx_model_H} is approximated for any user direction $\vec{\psi}_{\mathrm{r},1} = \vec{\psi}$ as follows
\begin{equation}\copyablelabel{eq:approximated_channel_los}
    r(\mathbf{m}, t) \approx g_{\mathbf{m}}(\vec{\psi}) \xi s(t) + \mathbf{f}_\mathrm{UE}^\mathrm{H}[k] \mathbf{n}(t),
\end{equation}
where $g_{\mathbf{m}}(\vec{\psi}) = \mathbf{a}^\mathrm{H}(\vec{\psi}) \boldsymbol{\Omega}(\mathbf{m}) \mathbf{a}(\boldsymbol{\psi}_{\mathrm{t},1})$ denotes the \gls{IRS} reflection gain for the virtual \gls{LoS} path. Furthermore, the combined influences of the \gls{LoS} component of the end-to-end channel, the \gls{BS} beamforming, and the user combining are represented by the scalar value
\begin{equation}
    \xi = \mathbf{f}_\mathrm{UE}^\mathrm{H}[k] \mathbf{a}(\boldsymbol\psi_\mathrm{UE,1}) [\mathbf{\Sigma}_\mathrm{r}]_{1,1} [\mathbf{\Sigma}_\mathrm{t}]_{1,1} \mathbf{a}^\mathrm{H}(\boldsymbol\psi_\mathrm{BS,1}) \mathbf{f}_\mathrm{BS}.
\end{equation}

Rewriting $\vec{\psi}$ using the notation introduced in Section~\ref{sec:system_model}, i.e., $\vec{\psi} = [\theta, \phi]^\mathrm{T}$, we notice that the right-hand side of \eqref{eq:approximated_channel_los} comprises three unknown scalar parameters, namely angles $\theta$ and $\phi$ and coefficient $\xi$.
Hence, for fixed values of $\theta$, $\phi$, and $\xi$, and even in the absence of noise, i.e., for $\vec{n}(t) = \mathbf{0}_{Q_\mathrm{UE}}$, at least three independent measurements are required in order to estimate $\theta$, $\phi$, and $\xi$.
To obtain such independent measurements, for our proposed direction estimation scheme, we assume that during each \gls{IDE} sub-block the \gls{IRS} is configured with different codewords from $\MIDE$, while the same pilot sequence of length $N$, $\mathbf{s} \in \mathbb{C}^{N}$, is repeatedly sent by the \gls{BS}.
For each considered codeword, we collect $N$ samples of the received signal, one for each pilot symbol, in vector $\vec{r}_{\vec{m}} \in \mathbb{C}^{N}$ and then utilize these $\vec{r}_{\vec{m}}$ to estimate the user's direction as detailed in the next subsection.
Denoting the set of codewords utilized for direction estimation in \gls{TB} $k$ by $\Mk \subseteq \MIDE$ and recalling that the symbol duration is denoted by $T_\mathrm{S}$, the duration of each \gls{IDE} sub-block for the proposed direction estimation scheme is given by $T_\mathrm{IDE} = |\mathcal{M}_k| N T_\mathrm{S}$.

%%%%%%%%%%%%%%%%%%%%%%%%%%%%%%%%%
\subsection{Direction Estimation}\copyablelabel{sec:direction_estimation}
To estimate the user's direction $\vec{\psi}$, we first recall that in addition to $\vec{\psi}$ also $\xi$ is unknown.
Furthermore, we assume that a proper direction estimation codebook (subset) for \gls{TB} $k$, $\Mk$, is available; we will detail later in Section~\ref{sec:user_tracking:codebook_and_grid_construction} how $\Mk$ can be constructed.
Then, the likelihood function for $\vec{\psi}$ and $\xi$ under the set of observations $\lbrace\vec{r}_{\vec{m}}\rbrace_{\vec{m}\in\Mk}$ is defined as
\begin{equation}\copyablelabel{eq:dir_est:likelihood}
    \mathcal{L}(\vec{\psi}, \xi) = \prod_{\mathbf{m} \in \Mk} f(\boldsymbol{r}_{\mathbf{m}} | \vec{\psi} , \xi),
\end{equation}
where
\begin{equation}\copyablelabel{eq:pdf_of_r}
\begin{split}
    & f(\mathbf{r}_{\mathbf{m}} | \boldsymbol{\psi}, \xi) = \frac{1}{\left(\pi Q_\mathrm{UE} \sigma^{2}\right)^N} \\
    & \times \exp\left( -\frac{1}{Q_\mathrm{UE} \sigma^2} \left \| \mathbf{r}_{\mathbf{m}} - g_{\mathbf{m}}(\boldsymbol{\psi} )\xi \mathbf{s}\right \|^2_2 \right),
\end{split}
\end{equation}
is the probability density function of a vector of complex Gaussian random variables with mean vector $g_{\mathbf{m}}(\vec{\psi} )\xi \mathbf{s}$ and covariance matrix $Q_\mathrm{UE} \sigma^2 \vec{I}_N$.

The joint estimation of $\vec{\psi}$ and $\xi$ according to \eqref{eq:dir_est:likelihood} results in a complicated non-convex problem that is in general infeasible to solve.
However, since our main goal is determining $\vec{\psi}$, an estimate for $\xi$ is actually not needed.
Hence, in order to reduce the computational complexity of estimating $\vec{\psi}$, we modify the likelihood function by eliminating the irrelevant {\em nuisance parameter} $\xi$ as \cite{reid2003likelihood}
\begin{equation}
    \tilde{\mathcal{L}}(\vec{\psi}) = \max_\xi \mathcal{L}(\vec{\psi}, \xi)
\end{equation}
and define the {\em peak \gls{ML} estimate} of $\vec{\psi}$, $\tilde{\boldsymbol\psi}$, as
\begin{equation}\copyablelabel{eq:dir_est:psi_tilde}
    \tilde{\boldsymbol\psi} = \arg\max_{\psi \in \mathcal{H}} \tilde{\mathcal{L}}(\vec{\psi}),
\end{equation}
where $\mathcal{H} \subseteq \mathbb{R}^2$ denotes the set of hypotheses to test for $\tilde{\boldsymbol\psi}$\footnote{For an overview on eliminating nuisance parameters see \cite{basu1977on}.}.

%%%%%%%%%%%%%%%%%%%%%%%%%%%%%%
In contrast to $\mathcal{L}$, $\tilde{\mathcal{L}}$ can be computed for any $\vec{\psi} \in \mathcal{H}$, since the following closed-form expression can be obtained
\begin{equation}\copyablelabel{eq:xi_problem}
	\tilde{\mathcal{L}}(\vec{\psi}) = \mathcal{L}(\vec{\psi}, \tilde{\xi}(\vec{\psi})),
\end{equation}
where
\begin{equation}\copyablelabel{eq:h_for_direction}
        \tilde{\xi}(\vec{\psi}) = \frac{\sum_{{\mathbf{m}}\in \Mk} g_{\mathbf{m}}^*(\vec{\psi}) \mathbf{s}^{\mathrm{H}} \mathbf{r}_{\mathbf{m}}}{\sum_{{\mathbf{m}}\in \Mk} N P_\mathrm{TX} | g_{\mathbf{m}}(\vec{\psi})|^2 },
\end{equation}
and $P_\mathrm{TX}$ denotes the transmit power at the \gls{BS}. Eqs.~\eqref{eq:xi_problem} and \eqref{eq:h_for_direction} follow directly from differentiating the logarithm of $\mathcal{L}(\vec{\psi}, \xi)$ with respect to $\xi$ and equating the resulting expression to zero.

Despite the availability of \eqref{eq:xi_problem} and \eqref{eq:h_for_direction}, the computational complexity of evaluating \eqref{eq:dir_est:psi_tilde} is still very high, since $\tilde{\mathcal{L}}(\vec{\psi})$ is in general not a convex function of $\vec{\psi}$. 
Consequently, a grid search is required to compute $\tilde{\boldsymbol\psi}$ in practice, i.e., $\mathcal{H}$ needs to be reduced to a finite set, and the accuracy of the grid search depends on the particular choice of $\mathcal{H}$.
Furthermore, depending on the expected \begin{change}direction\end{change} of the user in \gls{TB} $k$, i.e., depending on our {\em prior belief} about $\vec{\psi}$, different choices of $\mathcal{H}$ may be preferable.
Since this prior belief depends also on the estimated \begin{change}directions\end{change} of the user in previous \glspl{TB}, we allow $\mathcal{H}$ to vary from \gls{TB} to \gls{TB} and denote by
\begin{equation}\copyablelabel{eq:dir_est:psi_tilde_k}
    \tilde{\boldsymbol\psi}_k = \arg\max_{\boldsymbol\psi \in \mathcal{H}_k} \tilde{\mathcal{L}}(\vec{\psi}),
\end{equation}
the peak \gls{ML} estimate of $\vec\psi$ obtained under the set of hypotheses $\mathcal{H}_k$.
In the remainder of this section, we first make the notion of the aforementioned prior belief about $\vec\psi$ rigorous, before we specify $\mathcal{H}_k$ explicitly.

\begin{change}
\begin{remark}
\Copy{remark_gradient_ascent}{The direction estimation accuracy is lower bounded by the resolution of grid $\mathcal{H}_k$. In high \gls{SNR} scenarios, i.e., when the observations $\vec{r}_{\vec{m}}$ are strongly correlated with $\vec\psi$, a subsequent gradient ascent step starting from the direction $\tilde{\boldsymbol\psi}_k$ could be used to further increase the direction estimation accuracy beyond the grid resolution. However, this process is well established in the literature (e.g., \cite{teng2023variational}) and not considered here.}
\end{remark}
\end{change}

\subsection{Extrapolation of the User Direction}\label{sec:extrapolation_scheme}
In this subsection, we utilize the direction estimates $\tilde{\boldsymbol\psi}_{k}$ obtained in \glspl{TB} $k \leq k'$ to extrapolate the user's \begin{change}direction\end{change} to time interval $[t_{k'},t_{k'+1}]$, where we consider some fixed \gls{TB} $k' \geq 0$.
The extrapolated direction is then used to select a set of hypotheses $\mathcal{H}_{k'+1}$ and a codebook $\mathcal{M}_{k'+1}$ for direction estimation in \gls{TB} $k'+1$ according to Section~\ref{sec:direction_estimation}.
Furthermore, it is also utilized to configure the \gls{IRS} for \gls{CE} and \gls{D} in block $k'$ and for \gls{UC} in block $k'+1$.

\Copy{motivate_polynomial}{Since in practical scenarios users can exhibit very heterogeneous movement patterns, we propose the following generic $n$-th degree polynomial regression model for extrapolating the user's \begin{change}direction\end{change} at time $t \in [t_{k'},t_{k'+1}]$}
\begin{equation}\copyablelabel{eq:general_polynomial}
    \widehat{\boldsymbol\psi}_{k'}(t) = \begin{bmatrix}
\hat{\theta}_{k'}(t) \\
\hat{\phi}_{k'}(t)
\end{bmatrix} = \vec{B}_{k'} \begin{bmatrix}
    1 \\
    t \\
    \vdots\\
    t^n\\
\end{bmatrix},
\end{equation}
where $\widehat{\boldsymbol\psi}_{k'}(t)$ denotes the extrapolated user direction in $(\theta,\phi)$ coordinates and the first and second row of coefficient matrix $\vec{B}_{k'} \in \mathbb{R}^{2\times(n+1)}$ correspond to coefficient vectors $\vec{b}_{k'}^{(\theta)}$ and $\vec{b}_{k'}^{(\phi)}$, respectively.
Limiting the number of considered past direction estimates to $S_\mathrm{max}$ in order to neglect old, less correlated measurements, the coefficients $\vec{B}_{k'}$ minimizing the \gls{MSE} between the direction estimates $\tilde{\boldsymbol\psi}_{k}$ and the trajectory polynomial are obtained as
\begin{equation}\copyablelabel{eq:min_mse}
	\vec{B}_{k'} = \arg\min_{\vec{B}_{k'}'} \;\;\;   \frac{1}{S} \sum_{k=k'-S+1}^{k'} \left\| \tilde{\boldsymbol{\psi}}_{k} - \widehat{\boldsymbol{\psi}}(t_{k}) \right\|^2_2,
\end{equation}
where $S = \mathrm{min}\lbrace S_\mathrm{max},k'+1 \rbrace$. Since \eqref{eq:general_polynomial} is a linear function in $\vec{B}_{k'}$, the closed-form solution of \eqref{eq:min_mse} is readily obtained as \cite[Ch.\ 3]{demmel1997applied}
\begin{equation}\copyablelabel{eq:param_update}
\small
    \hspace{-2mm}
    \left(\vec{b}^{(\nu)}_{k'}\right)^{\mathrm{T}}
\hspace{-1mm} = \hspace{-1mm}
\begin{bmatrix}
\Pi_t(0) & \Pi_t(1) & \hspace{-2mm}\cdots\hspace{-2mm} & \Pi_t(n) \\
\Pi_t(1) & \Pi_t(2) & \hspace{-2mm}\cdots\hspace{-2mm} & \Pi_t(n+1) \\

\vdots & \vdots & \hspace{-2mm}\ddots\hspace{-2mm} & \cdots \\
\Pi_t(n) & \Pi_t(n+1) & \hspace{-2mm}\cdots\hspace{-2mm} & \Pi_t(2n) \\
 
\end{bmatrix}^{-1}
\begin{bmatrix}
\Pi_\nu(0)\\
\Pi_\nu(1)\\
\vdots \\
\Pi_\nu(n)
\end{bmatrix},\hspace{-2mm}
\end{equation}
where we have applied the definitions $\Pi_t(\rho) = \sum_{k=k'-S+1}^{k'} t_k^\rho$, $\Pi_\nu(\rho) = \sum_{k=k'-S+1}^{k'} \tilde{\nu}_k t_k^\rho$, for $\nu\in\{\theta,\phi\}$, and $\tilde{\boldsymbol\psi}_k = [\tilde{\theta}_k, \tilde{\phi}_k]^{\mathrm{T}}$.

Based on $\widehat{\boldsymbol\psi}_{k'}(t)$, for any codebooks $\MIDE, \MDT$, we are now able to determine the codeword that is best-aligned with the extrapolated \begin{change}direction\end{change} of the user at any time $t \in [t_{k'},t_{k'+1}]$.
In particular, we select
\begin{equation}\copyablelabel{eq:select_best_m_dt}
    \mathbf{m}^{\mathrm{DT}}_{k',\kappa} = \mathrm{arg}\min_{\mathbf{m} \in \MDT} \left \| \widehat{\boldsymbol{\psi}}_{k'}(t_{k',\kappa}) - \boldsymbol{\psi}_\mathrm{IRS}(\mathbf{m}) \right \|_2^2,
\end{equation}
$\kappa \in \lbrace 0,\ldots,\eta \rbrace$, to configure the \gls{IRS} at the beginning of each \gls{CE}, \gls{D}, and \gls{UC} sub-block in $[t_{k'},t_{k'+1}]$.

\begin{change}
\begin{remark}\label{remark_3}
    \Copy{ut_section_remark_configuration_complexity}{For the proposed trajectory extrapolation scheme, we need to choose the number of considered measurement points $S_\mathrm{max}$, as well as the order of the polynomial~$n$. Both these parameters have clear physical interpretations; $S_\mathrm{max}$ is directly proportional to the interval over which the autocorrelation of the user's movement is considerably high and $n$ is proportional to the number of turns that the user is expected to perform during this time. Hence, in practice, the parameters of the proposed scheme can be selected based on intuitive considerations. In Section \ref{sec:performance}, we consider a Kalman filter baseline scheme (see Fig.~\ref{fig:mse_over_time}). For the Kalman filter to achieve good accuracy, it is critical to choose the process noise and measurement noise models properly (for more details see Appendix~\ref{app:kalman_filter}), which can be very difficult in practice \cite{sharma2011alpha}. Hence, one advantage of the user tracking scheme presented in this section as compared to model-based tracking schemes like the Kalman filter is that the proposed scheme requires only the selection of two integer values, whereas, for example, the Kalman filter entails many more degrees of freedom. We will confirm in Section~\ref{sec:performance} that in terms of estimation accuracy the proposed scheme is comparable to the much more complicated Kalman filter.}
\end{remark}
\end{change}% end of ut_section_remark_configuration_complexity

\subsection{Codebook and Grid Construction}\label{sec:user_tracking:codebook_and_grid_construction}
In this section, we discuss how the extrapolated user trajectory can be utilized to properly select the direction estimation codebook $\mathcal{M}_{k'+1} \subseteq \MIDE$ and the set of hypothesis $\mathcal{H}_{k'+1}$ for direction estimation in time block $k'+1$, cf.~Section~\ref{sec:direction_estimation}.

To this end, we define
\begin{equation}\copyablelabel{eq:select_best_m_ide}
    \mathbf{m}^{\mathrm{IDE}}_{k'+1} = \argmin_{\mathbf{m} \in \MIDE} \left \| \widehat{\boldsymbol{\psi}}_{k'}(t_{k'+1}) - \boldsymbol{\psi}_\mathrm{IRS}(\mathbf{m}) \right \|_2^2,
\end{equation}
i.e., $\mathbf{m}^{\mathrm{IDE}}_{k'+1}$ denotes the direction estimation codeword whose main direction $\boldsymbol{\psi}_\mathrm{IRS}(\mathbf{m})$ is closest to the extrapolated user's direction at the beginning of the \gls{IDE} sub-block in \gls{TB} $k'+1$. Both $\Mkp$ and $\mathcal{H}_{k'+1}$ are constructed in the following from $\mathbf{m}^{\mathrm{IDE}}_{k'+1}$.

\paragraph{Construction of $\Mkp$}
Given that the trajectory extrapolation as described in the previous section is accurate and by its definition in \eqref{eq:select_best_m_ide}, codeword $\mideopt \in \MIDE$ is likely to yield a strong received signal for direction estimation in \gls{TB} $k'+1$.
We exploit this observation to further collect those codewords from $\MIDE$ for direction estimation whose beams most overlap with $\mideopt$, i.e., the codewords adjacent to $\mideopt$.
According to the construction of $\MIDE$ in Section~\ref{sec:codebook_model}, the set of the $(2\gamma + 1)^2 - 1$, $\gamma \in \lbrace 0, \ldots, \lfloor (M-1)/2 \rfloor \rbrace$ codewords adjacent to $\mideopt$ is given by
\begin{equation}
    \mathcal{M}^{(\gamma)}_{k'+1} = \{ \mathbf{m} \in \mathcal{M}^\mathrm{IDE} | \left\| \mathbf{m} - \mideopt \right\|_\infty \leq \gamma \},
\end{equation}
where $\gamma$ denotes the maximum difference between codeword indices.
Hence, the larger $\gamma$ is chosen, the more distinct measurements are available for direction estimation in \gls{TB} $k'+1$.
On the other hand, increasing $\gamma$ leads also to an increase in $\TIDE$, i.e., the overhead for direction estimation increases.
For the remainder of this paper, but without loss of generality, we set $\gamma=1$ and obtain $\Mkp = \mathcal{M}^{(1)}_{k'+1}$.

\begin{figure}
    \centering
    \includegraphics[clip, trim=5.3cm 16.6cm 3.8cm 4.5cm,width=0.49\textwidth]{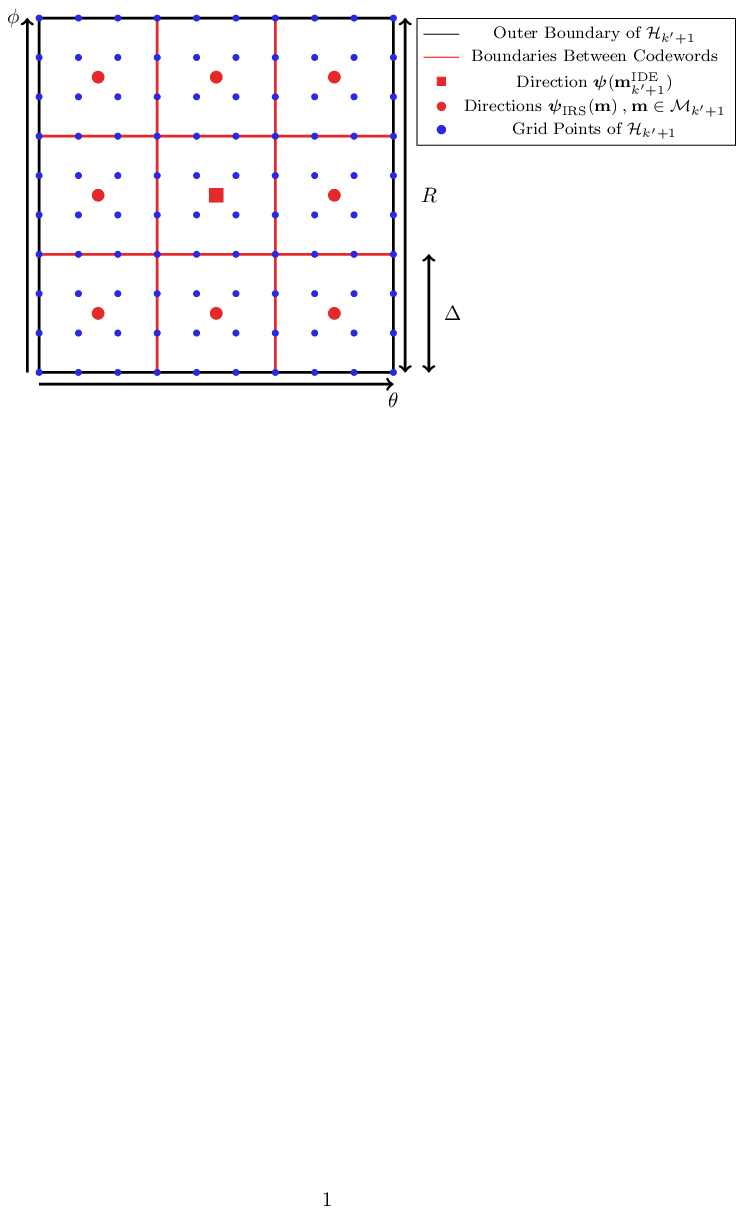}
    \vspace{-5mm}
	\caption{Visualization of the hypothesis grid $\mathcal{H}_{k'+1}$ with $H=10$.}
	\label{image:hypothesis_grid}
    \vspace{-3mm}
\end{figure}

According to this definition of $\Mkp$ and since $\MIDE$ is constructed by linearly shifting a predefined beam shape, cf.~Section~\ref{sec:codebook_model}, the main beam directions of the codewords $\vec{m} \in \Mkp$ relative to $\mideopt$ are arranged as depicted in Fig.~\ref{image:hypothesis_grid}.

%%%%%%%%%%%%%%%%%%%%%%%%%%%%%%%%%
\paragraph{Construction of $\mathcal{H}_{k'+1}$}

The angular spacing between the codewords in $\Mkp$ can be approximated from \eqref{eq:codeword_linear_shift} by evenly dividing the range of the phase difference between the array elements, $A_\mu(\boldsymbol\psi)\in[-1,+1], \mu\in\{1,2\}$, by the number of codewords $M$, i.e., $\Delta = \frac{2}{M} \frac{180^\circ}{\pi} [\mathrm{deg}]$, ignoring distortions for large angles. Since we consider $2\gamma+1=3$ adjacent codewords for each dimension, the total width of the range of directions that is covered by $\Mkp$ in each dimension is $R = 3\Delta$.
In order to span the hypothesis space $\mathcal{H}_{k'+1}$, we define equally spaced grid points within this range as
\begin{equation}
\small
    \begin{split}
        \theta_h & = \theta_\mathrm{IRS}(\mideopt) - \frac{R}{2} + R \frac{h-1}{H-1} , \\
        \phi_h & = \phi_\mathrm{IRS}(\mideopt) - \frac{R}{2} + R \frac{h-1}{H-1} ,
    \end{split}
\end{equation}
for $h\in\{1,...,H\}$.
Finally, $\mathcal{H}_{k'+1}$ is defined as $\mathcal{H}_{k'+1} = \lbrace (\theta_{h_1}, \phi_{h_2}) \,|\, h_1,h_2\in\{1,...,H\}  \rbrace$.

%%%%%%%%%%%%%%%%%%%%%%%%%%%%%%%%%%%%%%%%%%%%
\subsection{Integration of Direction Estimation and User Tracking into Time Block Structure}
Now, since all steps of the proposed direction estimation and user tracking algorithm have been introduced, we are ready to summarize the scheme and integrate it into the proposed time block structure.
To this end, we first note that in order to initialize the direction estimation in the first \gls{TB} according to Section~\ref{sec:direction_estimation}, the codebook $\mathcal{M}_0$ and the hypothesis set $\mathcal{H}_0$ are required.
Since for $k'=0$, no information about the user's \begin{change}direction\end{change} at $t_0$ is available from previous direction estimates, we assume that some initial access procedure is established to determine $\vec{m}^{\mathrm{IDE}}_0$ from which $\mathcal{M}_0$ and $\mathcal{H}_0$ are then constructed.
$\vec{m}^{\mathrm{IDE}}_0$ can, for example, be obtained by a full codebook search, see \cite{ning2020channel}.
Also, in the same manner, the initial codeword $\vec{m}^{\mathrm{DT}}_0$ can be obtained.

Based on the results from Sections~\ref{sec:direction_estimation} to \ref{sec:user_tracking:codebook_and_grid_construction} the proposed \gls{IRS}-assisted downlink communication scheme is summarized in Algorithm \ref{al:prediction_algorithm}.

\begin{algorithm}[]
\caption{Proposed User-tracking Scheme}
\begin{algorithmic}[1]
\small
    \STATE \textbf{Input:} $\vec{m}^{\mathrm{IDE}}_0$, $\vec{m}^{\mathrm{DT}}_0$
    \STATE \textbf{Initialize:} $k=0$
	\REPEAT
        \STATE Select \gls{IRS} codeword $\mathbf{m}^\mathrm{DT}_k$ according to \eqref{eq:select_best_m_dt} or $\mathbf{m}^\mathrm{DT}_0$ if $k=0$ (\gls{UC} sub-block).
        \STATE Determine user combining $\mathbf{f}_\mathrm{UE}[k]$ (\gls{UC} sub-block).
        \STATE Select $\mathbf{m}^\mathrm{IDE}_k$ according to \eqref{eq:select_best_m_ide}  or $\mathbf{m}^\mathrm{IDE}_0$ if $k=0$ and initialize $\mathcal{M}_k$ and $\mathcal{H}_k$ (\gls{IDE} sub-block).
		\STATE Estimate user direction as $\tilde{\boldsymbol\psi}_{k}$ according to \eqref{eq:dir_est:psi_tilde_k} (\gls{IDE} sub-block).
        \STATE Compute coefficient matrix $\vec{B}_k$ according to \eqref{eq:param_update} (\gls{IDE} sub-block).
        \STATE Set $\kappa=0$
		\STATE \textbf{repeat} $\eta$ \textbf{times}
            \STATE \;\;\;\; Extrapolate user direction as $\widehat{\vec{\psi}}_{k}(t_{k,\kappa})$ and select $\mathbf{m}^\mathrm{DT}_{k,\kappa}$ according to \eqref{eq:select_best_m_dt}.
	        \STATE \;\;\;\; Perform end-to-end channel estimation (\gls{CE} sub-block).
	        \STATE \;\;\;\; Transmit data (\gls{D} sub-block).
            \STATE \;\;\;\; $\kappa = \kappa + 1$
	    \STATE \textbf{end} \textbf{repeat}
     \STATE $k = k + 1$
\UNTIL{User exits the obstructed area.}
\end{algorithmic}\label{al:prediction_algorithm}
\end{algorithm}

%%%%%%%%%%%%%%%%%%%%%%%%%
%%%%%%%%%%%%%%%%%%%%%%%%%
%%%%%%%%%%%%%%%%%%%%%%%%%
\section{Direction Estimation Codebook Design}\copyablelabel{sec:codebook_generation}
\Copy{motivation_different_codebooks}{The direction estimation and user tracking scheme developed in the previous section is, in general, compatible with any choice of direction estimation codebook $\MIDE$. So, in principle, one could use the same codebook for direction estimation as for data transmission.
However, while the main requirement for the design of the data transmission codebook $\MDT$ is to maximize the reflection gain, i.e., the \gls{SNR}, the codebook requirements for direction estimation are different (as we will discuss further in this section) and, in general, it is beneficial to use two different codebooks for data transmission and direction estimation, i.e., $\MIDE \neq \MDT$.
Since no codebook design for \gls{IRS}-assisted direction estimation is available in the literature, in this paper, we propose a novel codebook design.}

To this end, we first propose an approximate theoretical model for the expected direction estimation error for a given codebook.
Then, we utilize this model to generate a novel codebook based on a modified beam shape that is specifically optimized for \gls{IRS}-assisted direction estimation.

%%%%%%%%%%%%%%%%%%%%%%%%%
\subsection{Direction Estimation Error Analysis}\label{sec:ide_codebook}

According to Section~\ref{sec:user_tracking:codebook_and_grid_construction}, the range of user directions to be estimated in \gls{TB} $k$ is $[\theta_{\mathrm{IRS}}(\vec{m}_k^{\mathrm{IDE}})-\frac{3}{2}\Delta, \theta_{\mathrm{IRS}}(\vec{m}_k^{\mathrm{IDE}})+\frac{3}{2}\Delta] \times [\phi_{\mathrm{IRS}}(\vec{m}_k^{\mathrm{IDE}})-\frac{3}{2}\Delta, \phi_{\mathrm{IRS}}(\vec{m}_k^{\mathrm{IDE}})+\frac{3}{2}\Delta] =: \bar{\mathcal{H}}_k$.
Since the direction estimation problem is identical for any $\vec{m}_k^{\mathrm{IDE}}$, we may assume without loss of generality $\theta_{\mathrm{IRS}}(\vec{m}_k^{\mathrm{IDE}}) = \phi_{\mathrm{IRS}}(\vec{m}_k^{\mathrm{IDE}}) = 0$.
Then, the expected squared direction estimation error\footnote{\Copy{CRB_not_feasible}{\begin{change} The \gls{MSE} realized with a given estimator is not necessarily identical to the corresponding \gls{CRB} and, hence, it is in general preferable to minimize the \gls{MSE} directly instead of the \gls{CRB} \cite{stoica1989music,stoica1990music}. However, which of the two approaches to select can in practice also depend on the mathematical and/or computational tractability of the respective problem formulation and optimization.\end{change}}
} for a given user \begin{change}direction\end{change} $\vec\psi$ can be written as
\begin{equation}\copyablelabel{eq:codebook_design:mse_v1}
    \mathrm{MSE}_k(\vec\psi) = \int\limits_{\bar{\mathcal{H}}_k} ||\vec\psi - \vec\psi' ||_2^2 \, p_{\tilde{\boldsymbol\psi}_k}(\vec\psi') \,\mathrm{d}\vec\psi',
\end{equation}
where $p_{\tilde{\boldsymbol\psi}_k}(\vec\psi') = \frac{\mathrm{d}}{\mathrm{d}\vec\psi'} \Pr \left\lbrace \tilde{\boldsymbol\psi}_k \leq \vec\psi' \right\rbrace$ and the peak \gls{ML} estimator $\tilde{\boldsymbol\psi}_k$ has been defined in Section~\ref{sec:direction_estimation}.
Since we make no prior assumptions with respect to where the user is located within the range of possible directions $\bar{\mathcal{H}}_k$, the expected mean squared direction estimation error in $\bar{\mathcal{H}}_k$ is given as
\begin{align}\copyablelabel{eq:codebook_design:mse}
    \mathrm{MSE}_k = \frac{1}{\left(3\Delta\right)^2} \int\limits_{\bar{\mathcal{H}}_k} \mathrm{MSE}_k(\vec\psi) \,\mathrm{d} \vec\psi.
\end{align}
Here, the challenge for computing \eqref{eq:codebook_design:mse} lies in evaluating the probability of {\em equivocation} $p_{\tilde{\boldsymbol\psi}_k}(\vec\psi')$, since $\tilde{\boldsymbol\psi}_k$ is a highly nonlinear function of the noisy direction estimation measurement signal $\vec{r}_{\vec{m}}$ as defined in Section~\ref{sec:tracking_scheme}.

\Copy{discussion_eq_26}{In order to develop a first-order approximation for $p_{\tilde{\boldsymbol\psi}_k}(\vec\psi')$, let us consider a single codeword $\vec{m}$ and a given measurement signal $\vec{r}_{\vec{m}}$ first.
Then, since $\vec{r}_{\vec{m}}$ is impaired by white Gaussian noise, cf.~\eqref{eq:pdf_of_r}, from all hypothesis directions $\vec\psi'$, the \gls{ML} estimate corresponds to the one with the smallest Euclidean distance $||\vec{r}_{\vec{m}}-g_{\vec{m}}(\vec\psi')\xi\vec{s}||_2$. \begin{change} Unfortunately, it is in general not clear how this Euclidean distance relates to $p_{\tilde{\boldsymbol\psi}_k}(\vec\psi')$, since the mapping of $\vec{r}_{\vec{m}}$ to $\tilde{\boldsymbol\psi}_k$ imposed by the \gls{ML} estimator is highly nonlinear and $p_{\tilde{\boldsymbol\psi}_k}(\vec\psi')$ can, hence, in general not be directly computed from the probability to observe $\vec{r}_{\vec{m}}$.\end{change}
However, since $g_{\vec{m}}(\vec\psi')$ is a continuous function in $\vec\psi'$ (according to the steering vectors $\vec{a}(\vec\psi')$ in Section~\ref{sec:system_model}), it can be approximated as a linear function in a neighborhood of $\vec\psi$, i.e., $g_{\vec{m}}(\vec\psi') = g_{\vec{m}}(\vec\psi) + g_{\vec{m}}'(\vec\psi)(\vec\psi - \vec\psi')$, where $g_{\vec{m}}'$ denotes the complex-valued derivative of $g_{\vec{m}}(\vec\psi)$ with respect to $\vec\psi$ and we have assumed that $\vec\psi'$ is close to $\vec\psi$.
In this case, since the complex-valued Gaussian noise in $\vec{r}_{\vec{m}}$ is circularly symmetric, it can be decomposed into a component parallel to $g_{\vec{m}}'(\vec\psi)$ and a second component that is orthogonal to $g_{\vec{m}}'(\vec\psi)$, where both components are zero-mean real-valued Gaussian random variables with variance $\sigma^2$.
Then, the \gls{ML} estimate of $\vec\psi$ is the projection of $\vec{r}_{\vec{m}}$ onto the line $c\, g_{\vec{m}}'(\vec\psi)$, $c \in \mathbb{R}$, i.e., the probability of equivocation between $\vec\psi$ and $\vec\psi'$ \begin{change} depends solely on the parallel noise component and is independent of the orthogonal noise component. 
In this case, the probability for obtaining a particular direction estimate $\vec\psi'$ is proportional to the probability of the corresponding parallel noise component, i.e.,\end{change}
\begin{equation}\copyablelabel{eq:26}
    p_{\tilde{\boldsymbol\psi}_k}(\vec\psi') \propto \exp\left(-\frac{||g_{\vec{m}}(\vec\psi)\xi\vec{s}-g_{\vec{m}}(\vec\psi')\xi\vec{s}||_2^2}{2 Q_{\mathrm{UE}}\sigma^2}\right),
\end{equation}
where $\propto$ indicates proportionality.}
Considering the general case when multiple codewords are available and applying the obtained first-order approximation to \eqref{eq:codebook_design:mse_v1}, we arrive at the following approximation
\begin{align}
    \mathrm{MSE}_k(\vec\psi) &\approx C_0 \int\limits_{\bar{\mathcal{H}}_k} ||\vec\psi - \vec\psi' ||_2^2\nonumber\\
    &\quad\prod_{\vec{m} \in \Mk} \exp\left(-\frac{||g_{\vec{m}}(\vec\psi)\xi\vec{s}-g_{\vec{m}}(\vec\psi')\xi\vec{s}||_2^2}{2 Q_{\mathrm{UE}}\sigma^2}\right)  \mathrm{d}\vec\psi'\nonumber\\ 
    &=: \widehat{\mathrm{MSE}}_k(\vec\psi),\copyablelabel{eq:27}
\end{align}
where $C_0$ is some positive scaling constant.
Accordingly, we define the following approximate mean squared direction estimation error for $\bar{\mathcal{H}}_k$
\begin{equation}\copyablelabel{eq:codebook_design:mse_approx}
    \widehat{\mathrm{MSE}}_k = C_1 \int\limits_{\bar{\mathcal{H}}_k} \widehat{\mathrm{MSE}}_k(\vec\psi) \,\mathrm{d} \vec\psi,
\end{equation}
where $C_1$ is a positive scaling constant that is irrelevant for the following optimization.
We will confirm in Section~\ref{sec:performance} that despite the approximate character of \eqref{eq:codebook_design:mse_approx}, it is indeed a useful criterion for practical direction estimation codebook design.

\subsection{Beam Shape Optimization}\label{sec:codebook_design:beam_shape_optimization}
In this section, we leverage the approximation of the direction estimation error derived in the previous section in order to optimize the \gls{IRS} beam shape $\omegaM$ for direction estimation.
The optimal beam shape minimizing $\widehat{\mathrm{MSE}}_k$, $\omegaM^*$, is defined as follows\footnote{Note that $\omegaM^*$ is independent of $k$ since the optimization for different values of $\bar{\mathcal{H}}_k$ differ only in terms of by how much $\omegaM^*$ is shifted, but not in terms of its optimal value.}
\begin{subequations}\copyablelabel{eq:codebook_design:omega_opt}
\begin{align}
    \omegaM^* &= \argmin_{\omegaM} \widehat{\mathrm{MSE}}_k,\\
    \mathrm{s.t.} &\quad |[\omegaM]_i|^2 = \bar{g}^2 \quad \forall \, i, \copyablelabel{eq:codebook_design:omega_opt_b}
\end{align}    
\end{subequations}
with the constant unit cell response factor $\bar{g}$.

Since $\widehat{\mathrm{MSE}}_k$ is not a convex function of $\omegaM$, \eqref{eq:codebook_design:omega_opt} cannot be solved efficiently.
However, \eqref{eq:codebook_design:omega_opt} can still be exploited to find a beam shape that is locally optimal in some neighbourhood of some existing beam shape $\omegaM^{(0)}$, where $\omegaM^{(0)}$ could for example be a linear or a quadratic phase shift profile \cite{jamali2020power}.
To this end, we write 
\begin{equation}\copyablelabel{eq:omega_is_exp_phase}
    \omegaM^{\mathrm{T}} = \bar{g} \cdot [\exp(\mathrm{j} \rho_1), \ldots, \exp(\mathrm{j} \rho_Q)],
\end{equation}
and for any beam shape $\omegaM$, we define the vector-valued phase shift function $\rho(\omegaM) = [\rho_1, \ldots, \rho_Q]^T \in \mathbb{R}^Q$ and its inverse function $\rho^{-1}(\cdot)$. Note that constraint \eqref{eq:codebook_design:omega_opt_b} is fulfilled automatically when using \eqref{eq:omega_is_exp_phase} to define $\omegaM$.

With these definitions, we equivalently reformulate \eqref{eq:codebook_design:omega_opt} in terms of $\rho(\omegaM)$ and define the locally optimal phase shift vector in the neighborhood of any given phase shifts $\rho\left(\omegaM^{(0)}\right)$ as
\begin{equation}\copyablelabel{eq:codebook_design:omega_opt_v2}
    \rho^*\left(\omegaM^{(0)}\right) = \argmin_{ \vec\rho_0 : \left|\left|\vec\rho_0-\rho\left(\omegaM^{(0)}\right)\right|\right|_2 < \delta} \widehat{\mathrm{MSE}}_k, 
\end{equation}
where $\vec\rho_0 \in \mathbb{R}^Q$, $\widehat{\mathrm{MSE}}_k$ depends on $\vec\rho_0$ via the beam pattern $\omegaM = \rho^{-1}(\vec\rho_0)$, and $\delta > 0$ is set small enough such that $\widehat{\mathrm{MSE}}_k$ is convex in the considered neighborhood of $\rho\left(\omegaM^{(0)}\right)$. Eq.~\eqref{eq:codebook_design:omega_opt_v2} can be solved efficiently as described in Appendix \ref{app:solution_of_eq:codebook_design:omega_opt_v2}. The locally optimal beam $\omegaM^{*}\left(\omegaM^{(0)}\right)$ is then obtained as $\omegaM^{*}\left(\omegaM^{(0)}\right) = \rho^{-1}\left(\cdot\right)$.
We will confirm in the following section that \eqref{eq:codebook_design:gradient_descent} indeed leads to a beam shape that is superior to existing beams in terms of the achieved direction estimation error.

%%%%%%%%%%%%%%%%%%%%%%%%%%%%%%%%
%%%%%%%%%%%%%%%%%%%%%%%%%%%%%%%%
%%%%%%%%%%%%%%%%%%%%%%%%%%%%%%%%
\section{Performance Evaluation}\copyablelabel{sec:performance}
In the following, the proposed direction estimation and \gls{UT} schemes are numerically evaluated.
First, in Subsection~\ref{sec:eval_direction_estimation}, the direction estimation codebook design presented in Section~\ref{sec:codebook_generation} is evaluated considering the direction estimation problem separately.
Then, in Subsection \ref{sec:eval_user_tracking}, the complete direction estimation and \gls{UT} scheme as introduced in Section~\ref{sec:tracking_scheme} is numerically evaluated. The simulation parameters used in this section are collected in Table \ref{table:simulation_parameters}.

\begin{table}[]
\centering
\caption{Simulation Settings. } \label{table:simulation_parameters}
\scalebox{0.75}{
\begin{tabular}{|cc||cc||cc|}
\hline
$\mathbf{p}_\mathrm{BS}$    & $[0,0,10]$ m   & $Q_{\mathrm{IRS},1}, Q_{\mathrm{IRS},2}$                   & 40, 40    & $T$ & 1.5 s                       \\ \hline
$\mathbf{p}_\mathrm{IRS}$   & $[-40,40,5]$ m & $d_1, d_2$                   & $\lambda/2$ & $T_\mathrm{CE}+T_\mathrm{D}$                  &   1.29 ms                   \\ \hline
$\mathbf{r}_\mathrm{C}$ & $[0,40,0]$ m  & $Q_\mathrm{BS}, Q_\mathrm{UE}$   & 16x4 , 2x2  & $T_\mathrm{S}$                & 4.16 $\mu$s \\ \hline
$r_1, r_2$                          &   10 m, 5 m     & $L_\mathrm{r}$, $L_\mathrm{t}$ & 4, 4        & $K_\mathrm{t}, K_\mathrm{r}$                       &   10,10                   \\ \hline
$v$           &    5 km/h        &    $S_\mathrm{max}$                          &  3         &      $\sigma^2$                &       - 120 dBm            \\ \hline
$N_\mathrm{IDE}$        & 5         &   $n, \gamma$                &    1,1  &   $f_\mathrm{c}$         &    28 GHz                     \\ \hline
\end{tabular}
}
\vspace{-3mm}
\end{table}

\subsection{Direction Estimation}\label{sec:eval_direction_estimation}
In this section, we first present the beam shape generated by the local optimization procedure from Section~\ref{sec:codebook_design:beam_shape_optimization}.
Then, the direction estimation performance of the codebook obtained from this beam is evaluated and compared to a benchmark codebook from the literature.

\subsubsection{Beam Shape}\label{sec:example_realization}

\Copy{fig:beamshape_abs}{
\begin{figure}
        \centering
        \includegraphics[width=0.45\textwidth]{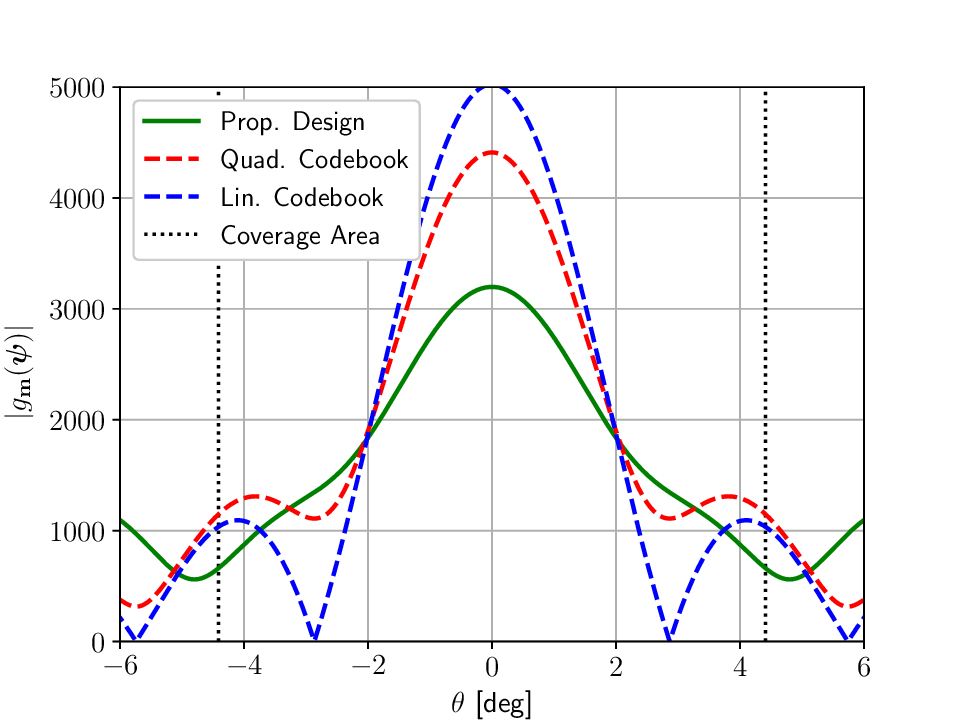}
        \vspace{-2mm}
        \copyablefigurecaptionpluslabel{The beam shapes for the proposed direction estimation codebook design (green), the quadratic codebook (red), and the linear codebook (blue) for different horizontal user directions, i.e., different values of $\theta$. Furthermore, the desired coverage area is indicated (dotted vertical lines).}{fig:beams_amplitude}
        \vspace{-3mm}
\end{figure}
}

Fig.~\ref{fig:beams_amplitude} shows the beam amplitudes $|g_{\vec{m}}(\vec\psi)|$ for different horizontal directions $\theta$ as generated by the linear codebook, the quadratic codebook \cite{jamali2020power}, and the direction estimation codebook design presented in this paper, all for an \gls{IRS} of size $Q=40$ and \begin{change}$M=25$\end{change}. For the proposed design, the quadratic phase shift profile, that is also given as benchmark, is used as the initial phase shift profile $\boldsymbol{\omega}_\mathcal{M}^{(0)}$. \Copy{define_quad_beamshape}{\begin{change} The quadaratic phase shift profile is defined as \cite{jamali2020power}
\begin{equation}\copyablelabel{eq:quadratic_phase_shifts}
    \rho_q = - \frac{2\pi d_\mathrm{IRS}}{M \lambda} \left( \frac{\Delta\beta_m}{2 Q} + \beta_m q \right), q\in\{1,...,Q\},
\end{equation}
where $m\in\{{0,...,M-1}\}$, $\Delta\beta = \frac{\bar{\beta}}{M}$, $\beta_m = m \Delta\beta$, and $\bar\beta = \min\{4,\lambda/d_\mathrm{IRS}\}$.
\end{change}}
We observe from Fig.~\ref{fig:beams_amplitude} that the direction estimation beam shape proposed in this paper, i.e., $\omegaM^*(\omegaM^{(0)})$ as defined in Section~\ref{sec:codebook_design:beam_shape_optimization}, is broader as compared to the linear and quadratic reference beams.
The broad beam shape is intuitively advantageous for direction estimation, since the entire considered direction range is supplied with sufficient power to accurately estimate the user's actual \begin{change}direction\end{change}. In contrast, the considered reference beams are optimized for data transmission which benefits from a concentrated high \gls{SNR} at the center of the beam. \Copy{beamshape_description_1}{\begin{change}An important observation is that, unlike both reference designs, the proposed design does not have sidelobes within the considered range. Intuitively, this avoids scenarios, where several adjacent angles $\theta$ yield a similar reflection gain $g_{\vec{m}}(\theta)$, which results in similar reflection gain hypotheses that are likely to be confused by the direction estimation algorithm.\end{change} Finally, we observe from Fig.~\ref{fig:beams_amplitude} that the maximum gain of the proposed beam is lower than the maximum gains of the reference beams.}
However, we will confirm in the following section that the negative effect of the smaller amplitude is more than compensated for by the advantages of the proposed beam design in terms of the direction estimation performance.

\subsubsection{Performance}

Different codebooks may lead to different reflection gains and therefore different \glspl{SNR} at the receiver, even if all other system parameters are identical.
Hence, to compare the direction estimation performances of different codebooks, we first define the {\em \gls{MSNR}} as follows
\begin{equation}
    \mathrm{MSNR} = \frac{\xi^2 g_\mathrm{max}^2 P_\mathrm{TX} }{\sigma^2},
\end{equation}
where $g_\mathrm{max} = \bar{g} Q^2$.
The \gls{MSNR} is a codebook independent quantity corresponding to the maximum achievable \gls{SNR} at the receiver for the optimal \gls{IRS} phase shift configuration.
Then, for several randomly generated test directions $\vec\psi$ within the boundaries of some hypothesis set $\mathcal{H}_k$, i.e., $\vec\psi \in \bar{\mathcal{H}}_k$, and different codebooks, the \gls{IDE} as presented in Section~\ref{sec:tracking_scheme} is performed and the squared error $\left\| \tilde{\vec{\psi}}_k - \vec{\psi} \right\|_2^2$ is evaluated.

\Copy{fig_mse_vs_msnr_with_music}{
\begin{change}
\begin{figure}
    \centering
    \vspace{-5mm}
    \includegraphics[width=0.45\textwidth]{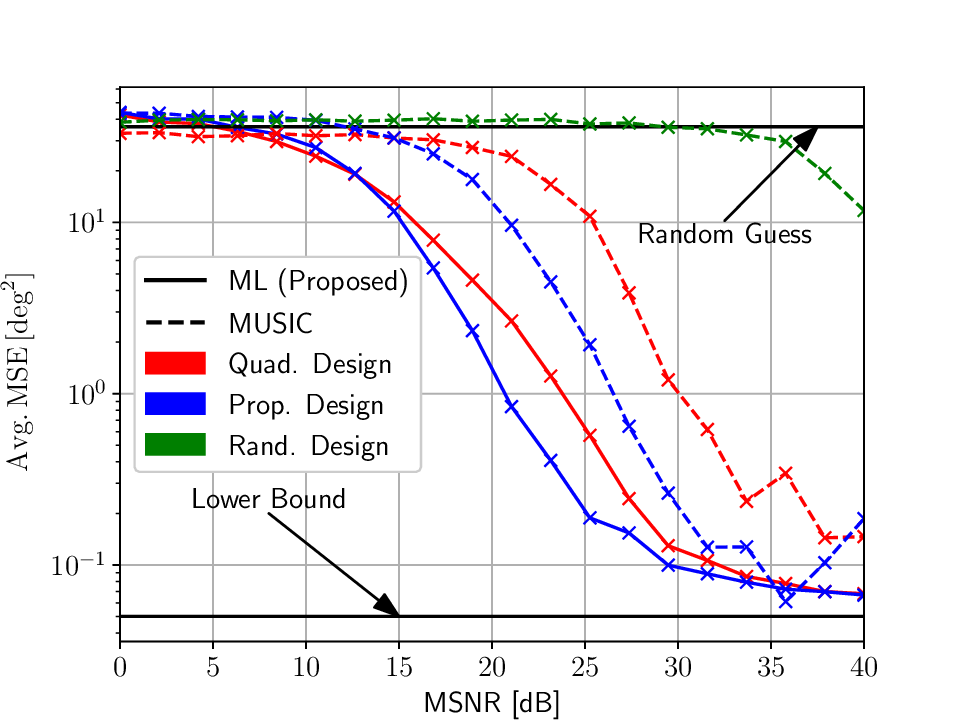}
    \vspace{-2mm}
    \copyablefigurecaptionpluslabel{Average direction estimation error for the quadratic (red), the proposed direction estimation codebook (blue), and random phase shift configurations (green) for different \glspl{MSNR}. The proposed \gls{ML}-based approach is compared to the \gls{MUSIC} algorithm.}{fig:beams_performance}
    \vspace{-5mm}
\end{figure}
\end{change}}

\Copy{evaluation_ml_vs_music}{Fig.~\ref{fig:beams_performance} shows the average direction estimation error obtained for the quadratic reference codebook and the direction estimation codebook proposed in this paper, respectively\footnote{Using the linear codebook results in large intervals between the narrow main lobes of the codewords, where the direction estimation performs so poor that its average performance would be difficult to illustrate in one plot with the two considered codebooks. Hence, the linear codebook is not considered in Fig.~\ref{fig:beams_performance}.}.
\begin{change} Furthermore, we compare the proposed direction estimation approach from Section~\ref{sec:direction_estimation} to the \gls{MUSIC} estimation algorithm from \cite{wang2023music} for random \gls{IRS} phase shifts, as proposed in \cite{wang2023music}, as well as for the two considered codebooks. A detailed description of the \gls{MUSIC} algorithm is provided in Appendix \ref{app:music_algorithm}.\end{change}
Additionally, the \gls{MSE} for a \textit{random guess} of a point in $\mathcal{H}$ and the \textit{lower bound} \gls{MSE}, when always selecting the closest hypothesis direction from $\mathcal{H}_k$, are plotted. We note that since for very low \glspl{MSNR}, the presented direction estimation method tends to mistake the far left and far right sections (with respect to the coverage area) of the considered beams, due to their similar reflection gains, the resulting \gls{MSE} may be worse than an unbiased random guess in this case.

We observe from Fig.~\ref{fig:beams_performance} that for increasing \glspl{MSNR} the \gls{MSE} decreases.
This is intuitive, since the noise becomes less detrimental as the \gls{MSNR} increases. 
\begin{change}Furthermore, we observe that the \gls{MUSIC} algorithm with random \gls{IRS} phase shifts performs worst. The main reason for this is that the reflection beams for random \gls{IRS} phase shifts are misaligned with the user, resulting in low \gls{SNR} at the receiver. \Copy{codebook_good_for_DE}{Also, the proposed direction estimation codebook outperforms the quadratic reference codebook over a wide range of \glspl{MSNR} for both considered direction estimation schemes (\gls{ML} and \gls{MUSIC}),} while both direction estimation schemes approach the \textit{random guess} and the \textit{lower bound} for low and high \glspl{MSNR}, respectively. For all considered codebooks, the proposed \gls{ML}-based scheme outperforms the \gls{MUSIC} algorithm. 
However, the \gls{MUSIC} algorithm may have a lower computational complexity in practice \cite{stoica1989music}\footnote{\begin{change}For more details on the relationship between \gls{ML} and \gls{MUSIC} estimators, see Section VI in \cite{stoica1989music}.\end{change}}.
Therefore, in scenarios where the computational complexity is strictly limited, the \gls{MUSIC} algorithm may be a viable alternative.\end{change}
Fig.~\ref{fig:beams_performance} confirms that the direction estimation beam shape design presented in Section~\ref{sec:codebook_generation} is superior to existing reference beam designs in terms of direction estimation performance.}
% \cite{stoica1989music} VI. THE RELATIONSHIP BETWEEN THE MUSIC AND ML ESTIMATOR
% \cite{stoica1989music} When the n signals are uncorrelated, it should indeed be possible (at least for N) to decouple the n-dimensional search problem implied by the MLE into the n one-dimensional problems solved by MUSIC. When the signals are correlated, this should not be possible. Note that it is this decoupling that makes the MUSIC estimator much more attractive computationally than the MLE.
% \cite{stoica1990music} In this light, comparison between the MLE and the (much) simpler computationally MUSIC estimator becomes of significant interest.
% \cite{stoica1990music} Computationally, MUSIC is such a simpler estimator but is seldomly more accurate than MLE
%%%%%%%%%%%%%%%%%%%%%%%%%%%%%%%%%%%%%%%%%%%%%%%%%%%%%%
\subsection{Performance of \gls{UT} Scheme}\label{sec:eval_user_tracking}
In this section, we compare the average realized \gls{SNR} during data transmission and the effective rate obtained with the proposed \gls{UT} algorithm, cf.~Section~\ref{sec:tracking_scheme}, and three baseline schemes. For data transmission and direction estimation, we adopt the quadratic codebook design and the proposed direction estimation codebook design, respectively.

Specifically, the \gls{SNR} is evaluated as follows
\begin{equation}
    \mathrm{SNR}(t) = \frac{|\mathbf{f}_\mathrm{UE}^\mathrm{H}[k] \mathbf{H}_\mathrm{r} \mathbf{\Omega}(\mathbf{m}^\mathrm{DT}) \mathbf{H}_\mathrm{t} \mathbf{f}_\mathrm{BS} s(t)|^2}{Q_\mathrm{UE}\sigma^2},
\end{equation}
and the effective rate is defined as
\begin{equation}\label{eq:evaluation:def_gamma}
    R(t)=(1-\Gamma) \log_2(1+\mathrm{SNR}(t)),
\end{equation}
where $\Gamma$ denotes the overhead ratio.
The specific value of $\Gamma$ depends on the employed transmission scheme as will be discussed below.

\Copy{explain_codebook_search}{The following three baseline schemes are considered. \begin{change}First, the \textit{focusing baseline} has perfect \gls{CSI} knowledge and aligns the reflected beam to the \gls{LoS} direction at all times. 
Second, we consider a \textit{perfect codeword selection} that employs the codeword resulting in the highest received power at all times without any delay.
Third, the \textit{hierarchical codebook search} operates in two steps during each \gls{IDE} block. In the initial step, a codebook with few, but wide codewords is considered, where each codeword covers a relatively large area. The \gls{BS} repeatedly sends a pilot sequence while the \gls{IRS} cycles through all available codewords in each \gls{IDE} block.
Subsequently, a codebook with more but narrower beams is utilized, where each codeword covers a subspace of the codewords from the previous step's codebook, focusing only on the beams corresponding to the previously determined approximate direction. In both steps, the codeword that results in the highest received power is selected.
A possible practical implementation of the \textit{hierarchical codebook search} is presented in \cite{liu2024hierarchical}.\end{change}
\Copy{explain_loss_probability}{\begin{change}To obtain upper bounds on the performance of the considered baseline schemes, we consider "perfect" versions that optimally select the codeword that provides the largest reflection gain. Therefore, all baseline schemes track the correct (approximate) direction of the user at all times.\end{change}}
} % end of explain_codebook_search
For all baselines, the \gls{BS} and user antenna beamforming are carried out as for the proposed \gls{UT} scheme.
The results have been averaged over 12 different movement trajectories, each simulated 10 times to account for different noise realizations. \Copy{eval_num_samples_trajectory}{\begin{change}For each trajectory, we collect $3000$ random noise samples, resulting in a representative total number of data points considered in the evaluation.\end{change}}
Simulations in which a permanent user connection could not be maintained by the system were excluded from the evaluation, since those cases would be dealt with on a higher protocol layer than the physical layer in any real-world application.
However, the percentage of excluded simulations, denoted as \textit{loss probability}, is reported for all simulations. The quadratic codebook design is used for all reference schemes and for $\MDT$ and the proposed codebook design method from Section~\ref{sec:codebook_generation} is used to generate $\MIDE$.

%%%%%%%%%%
\subsubsection{Overhead}\label{sec:overhead}
Next, we classify the overhead ratio $\Gamma$ for each of the considered schemes before we discuss the numerical results.
\paragraph{Proposed UT Scheme}
We recall from Section~\ref{sec:system_model} that the duration of the \gls{CE} sub-block is $T_\mathrm{CE} = N_\mathrm{CE} T_\mathrm{S}$, where $N_\mathrm{CE}$ is the number of pilot symbols for the end-to-end channel estimation.
Since the parameter estimated in the \gls{CE} block, $\xi$, is a scalar that does not scale with the number of \gls{IRS} unit cells, the required number of pilot symbols for channel estimation is constant, e.g., $N_\mathrm{CE} \sim \mathcal{O}(1)$, resulting in a short $T_\mathrm{CE}$.
Overall, for the proposed \gls{UT} scheme,
\begin{equation}
    \Gamma = \frac{T_\mathrm{UC} + T_\mathrm{IDE} + \eta T_\mathrm{CE}}{T}.
\end{equation} 

%%%%%%%%%%
\paragraph{Focusing Baseline}
The \gls{TB} structure of the focusing baseline scheme consists of alternating \gls{CE} and \gls{D} sub-blocks, resulting in a relative overhead of:
\begin{equation}
    \Gamma_\mathrm{B} = \frac{T_\mathrm{CE}^\mathrm{B}}{T_\mathrm{CE}^\mathrm{B} + T_\mathrm{DT}^\mathrm{B}},
\end{equation}
where $T_\mathrm{CE}^\mathrm{B} = N_\mathrm{CE}^\mathrm{B} T_\mathrm{S}$ with pilot sequence length $N_\mathrm{CE}^\mathrm{B}$.
For this baseline, the individual channel of each \gls{IRS} unit cell ($\mathbf{H}_\mathrm{r}$ and $\mathbf{H}_\mathrm{t}$), from now on called the full channel, needs to be estimated.
The number of unknown channel parameters of the full channel scales with the number of \gls{IRS} unit cells and conventional channel estimation would require the pilot sequence length to scale with the number of unknown channel parameters. As this is too resource expensive, and potentially infeasible to perform within the channel coherence time, \gls{CS} methods can be used. We do not analyze compressed sensing methods in this paper, but from \cite{wang2020compressed} it can be concluded that the required number of pilot symbols scales as $N_\mathrm{CE}^\mathrm{B} \sim \mathcal{O} \left ( L_\mathrm{r} L_\mathrm{t} \mathrm{ln}( Q_\mathrm{IRS})\right)$.

%%%%%%%%%%
\paragraph{Hierarchical Search Baseline}
For the hierarchical search baseline, we assume that there are $L_\mathrm{C}$ codebook levels and each codeword in a lower level covers the range of $N_\mathrm{HS}$ codewords in the next level. In the first codebook level, all $M_\mathrm{HS}/N_\mathrm{HS}^{L_\mathrm{C}-1}$ codewords are considered, where $M_\mathrm{HS}$ is the size of the final codebook level. In all $L_\mathrm{C}-1$ remaining codebook levels, $N_\mathrm{HS}$ codewords are considered, if only the best result of the previous level is selected.
In summary, the overhead time is given as $T_\mathrm{HS}= T_\mathrm{CE} \big(M_\mathrm{HS}/N_\mathrm{HS}^{L_\mathrm{C}-1} + N_\mathrm{HS} (L_\mathrm{C}-1)\big)$ and, hence, 
\begin{equation}
    \Gamma_\mathrm{HS} = \frac{T_\mathrm{UC} + T_\mathrm{HS} + \eta_\mathrm{HS} T_\mathrm{CE}}{T} ,
\end{equation}
where $\eta_\mathrm{HS}$ is the number of \gls{CE} and \gls{D} sub-blocks for the hierarchical search baseline.

%%%%%%%%%%
\paragraph{Perfect Codeword Selection Baseline}
The overhead and, hence, the effective rate of the perfect codeword selection cannot be determined exactly, since it is a purely theoretical bound with potentially infinite overhead. However, for comparison, in the following we utilize an approximation of the overhead incurred in the perfect codeword selection based on its scaling order.\\

The exact numerical value of the overhead for each considered scheme depends on many factors in the system model and employed algorithms, which makes it infeasible to obtain. 
Hence, for the following numerical evaluation, the length of the pilot symbol sequences are approximated by their scaling orders, i.e., $N_\mathrm{CE} = 1$ for the proposed \gls{UT} scheme, the perfect hierarchical search, and the perfect codeword selection, and $N_\mathrm{CE}^\mathrm{B} = L_\mathrm{r} L_\mathrm{t} \mathrm{ln}(Q_\mathrm{IRS} )$ for the full \gls{CSI} estimation in the focusing baseline.
For the hierarchical search, we assume a codebook depth of $L_\mathrm{C} = 2$ and that each codeword in the lower codebook level covers $N_\mathrm{HS} = 4$ codeword in the succeeding codebook level. 

%%%%%%%%%%%%%%%%%%%%%%%%%%%%%%%%%%%%%
\subsubsection{User Tracking Evaluation}
For the following evaluation, several possible movement trajectories, linear as well as nonlinear movements, were simulated.

%%%%%%%%%%%%%%%%%%%%%%%%%%%
\paragraph{Linear Movement}
We first consider linear movement, which occurs in many practical scenarios, such as a person traveling along a straight street.
Fig.~\ref{fig:linear_snr_rate_power}(a) illustrates the average \gls{SNR} during data transmission for the proposed \gls{UT} scheme and the considered baseline schemes\footnote{The range of considered transmit powers is relatively large to show the asymptotic behaviour for error-free \gls{UT} in high \gls{SNR} scenarios.}.
For the proposed \gls{UT} scheme, the perfect hierarchical search, and the perfect codeword selection two different codebook sizes are considered.
As expected, the focusing baseline achieves the highest \gls{SNR} among all considered schemes.
This result is intuitive, since the IRS reflection gain is maximized towards the actual \gls{LoS} user direction in the focusing baseline.
For all other schemes, the realizable reflection gain depends on the codebook size. Naturally, a larger codebook enables narrower beams with a larger reflection gain.
Consequently, we observe from Fig.~\ref{fig:linear_snr_rate_power} that the codebooks of size $M = 40$ achieve a higher \gls{SNR} than those with codebook size $M = 30$.
For both codebook sizes, the proposed \gls{UT} scheme performs very close to the optimum, i.e., the perfect codeword selection, while the performance of the hierarchical search baseline is comparatively worse.
This observation is readily explained by the fact that the codeword selection in the hierarchical search baseline is quickly outdated since it is performed only once towards the end of each \gls{TB}.
Furthermore, it can be observed that the loss probability for the proposed \gls{UT} scheme goes to zero for all but very small transmit powers. Hence, a recovery procedure is not needed to track linear movements.

Fig.~\ref{fig:linear_snr_rate_power}(b) shows the average effective rate for all considered schemes for different transmit powers.
We observe from Fig.~\ref{fig:linear_snr_rate_power}(b) that, due to its low overhead, the proposed \gls{UT} scheme outperforms all considered baseline schemes in terms of the average effective rate.

The observations in this section confirm that the key features of the proposed \gls{UT} scheme, namely the user \begin{change}direction\end{change} extrapolation algorithm presented in Section~\ref{sec:tracking_scheme} and the direction estimation codebook design from Section~\ref{sec:codebook_generation}, indeed lead to competitive performance of the proposed scheme when compared to baseline schemes.

\begin{figure}
     \centering
     \begin{subfigure}{0.45\textwidth}
        \includegraphics[clip, trim=0cm 0cm 0cm 12mm,width=\textwidth]{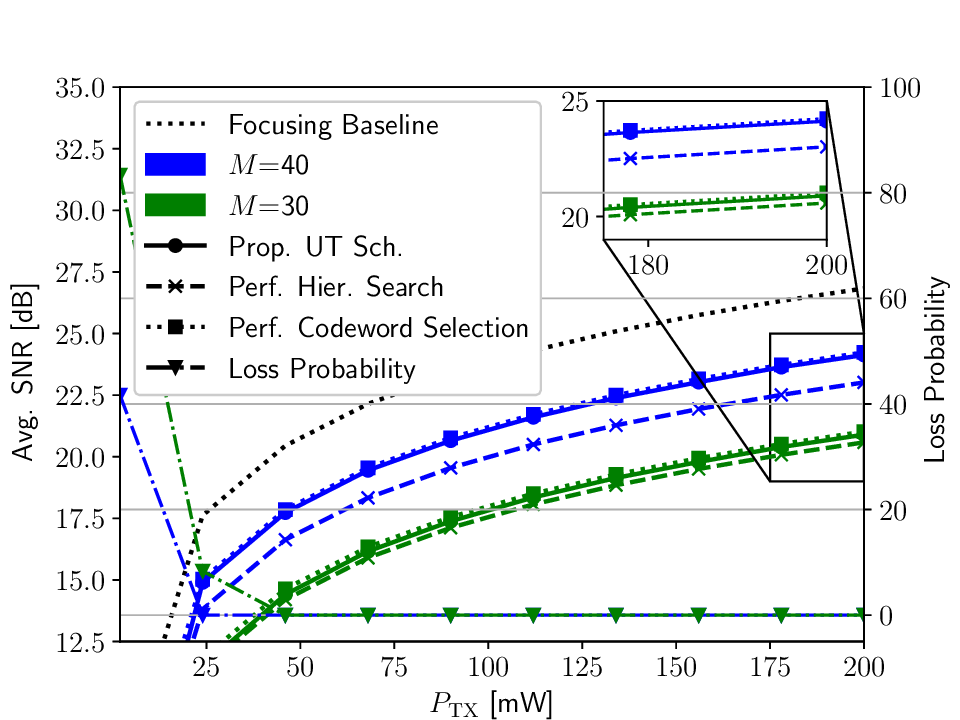}
        \caption{}
    \end{subfigure}
    \hfill
    \begin{subfigure}{0.45\textwidth}
        \includegraphics[clip, trim=0cm 0cm 0cm 12mm,width=\textwidth]{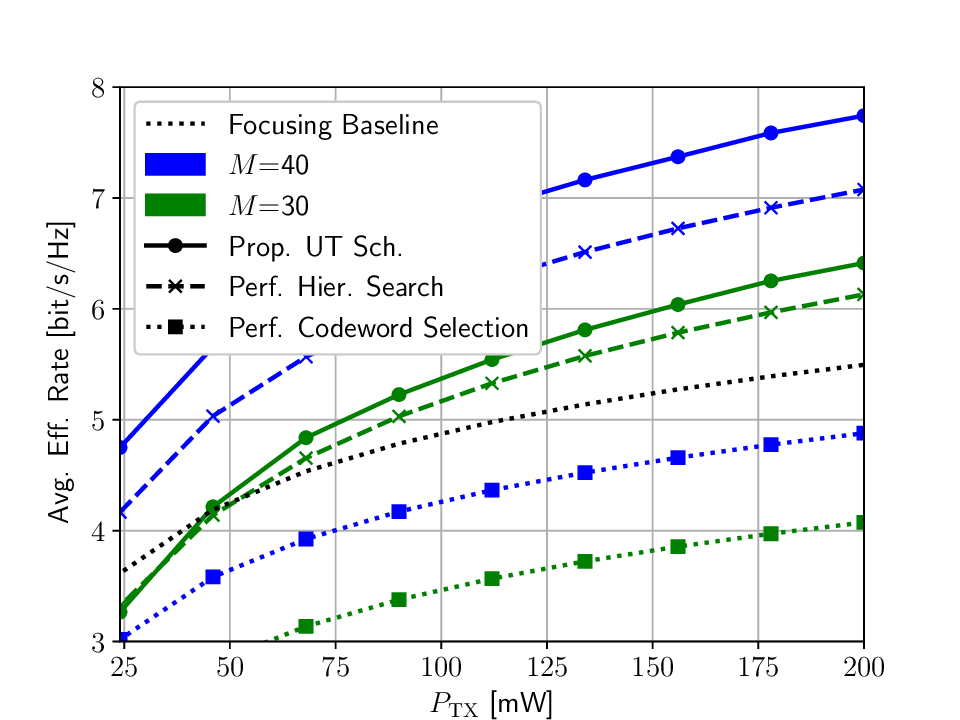}
        \caption{}\label{fig:linear_rate_vs_power}
    \end{subfigure}
    \vspace{-2mm}
    \copyablefigurecaptionpluslabel{Performance averaged over several iterations of linear movement. (a) Average \gls{SNR}. (b) Average effective rate.}{fig:linear_snr_rate_power}
    \vspace{-5mm}
\end{figure}

%%%%%%%%%%%%%%%%%%%%%%%%%%%%%%%
\paragraph{Nonlinear Movement}
To explore the limits of the proposed \gls{UT} scheme, we consider now a highly nonlinear movement trajectory, similar to the one introduced in \cite{dehkordi2021adaptive}.
Specifically, the user is assumed to move inside a circular area with radius $r_1$ and center $\mathbf{r}_\mathrm{C}$. Starting at a random point on the outer boundary of the blocked area, the user moves linearly towards the center until it reaches a distance of $r_2$, $r_2 < r_1$, to the center. Then, it follows a circular trajectory around the center in a counter-clockwise direction. At a randomly determined \begin{change}direction\end{change}, the user again adopts a linear trajectory away from the center until leaving the blocked area.
Hence, the user trajectory comprises first a linear, then a nonlinear, and then a second linear part, with abrupt (non-smooth) direction changes between the individual parts.

In Fig. \ref{fig:snr_rate_power}(a), the average \gls{SNR} for nonlinear movement is presented, with otherwise the same simulation setup as for the linear movement evaluated in Fig. \ref{fig:linear_snr_rate_power}. 
We observe from Fig.~\ref{fig:snr_rate_power}(a) that the average \gls{SNR} obtained with the proposed \gls{UT} scheme is still slightly larger than that of the hierarchical search baseline.
On the other hand, the gap in terms of \gls{SNR} to the perfect codeword selection is increased as compared to the linear movement.
In general, a large codebook with narrow beams requires more frequent changes of the employed codeword. Hence, the gain obtained by employing user \begin{change}direction\end{change} extrapolation in the proposed scheme as compared to the perfect hierarchical search baseline is greater for $M$=$40$ than $M$=$30$.

\begin{figure}
    \centering
    \begin{subfigure}{0.45\textwidth}
        \includegraphics[clip, trim=0cm 0cm 0cm 12mm,width=\textwidth]{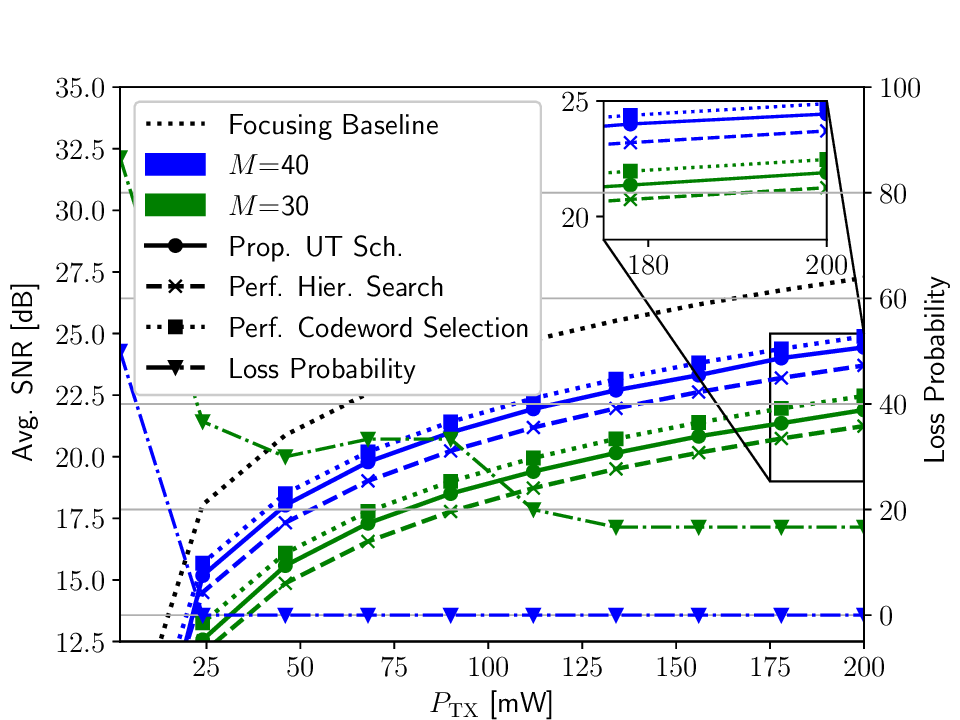}
        \caption{}\label{fig:snr_rate_power_a}
    \end{subfigure}
    \hfill
    \begin{subfigure}{0.45\textwidth}
        \includegraphics[clip, trim=0cm 0cm 0cm 12mm,width=\textwidth]{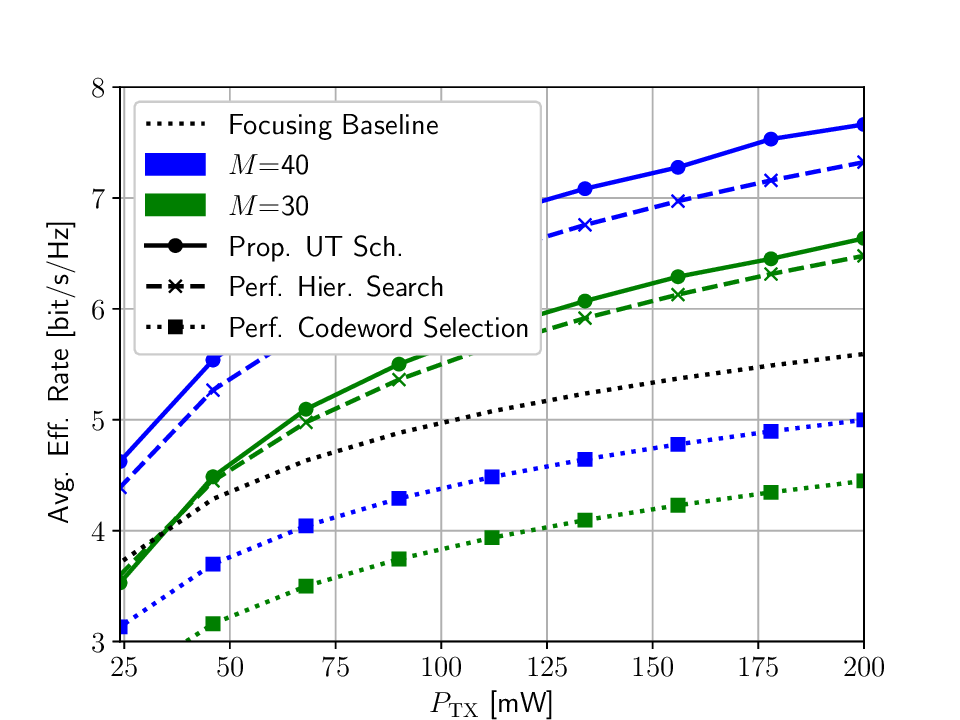}
        \caption{}\label{fig:snr_rate_power_b}
    \end{subfigure}
    \vspace{-2mm}
    \copyablefigurecaptionpluslabel{Performance averaged over several iterations of nonlinear movement. (a) Average \gls{SNR}. (b) Average effective rate.}{fig:snr_rate_power}
    \vspace{-5mm}
\end{figure}

The \textit{loss probability} of the \gls{UT} algorithm employing the larger codebook is zero for most values of the transmit power, similar to the linear movement case.
However, for the smaller codebook, a non-negligible loss probability persists even for large transmit powers.
Since the tracking loss of a user is a multi-factorial event, it is in general not possible to attribute it to any single event.
However, since no tracking losses occur for the linear movement, some losses in the nonlinear scenario can be attributed to the aforementioned challenges in predicting the nonlinear user trajectory, especially when the user direction changes abruptly, i.e., when the trajectory is highly non-smooth.
We will confirm this hypothesis towards the end of this section in the discussion of Fig.~\ref{fig:mse_over_time}.

Fig.~\ref{fig:snr_rate_power}(b) shows the average effective rate for the considered nonlinear movement as a function of the transmit power.
We observe from Fig.~\ref{fig:snr_rate_power}(b) that the focusing baseline performs worse compared to the other considered schemes, similar to the linear movement case.
\begin{change}For both codebook sizes, the proposed \gls{UT} scheme still achieves a higher effective rate than the hierarchical search baseline, but the gap is slightly decreased as compared to the linear movement.\end{change}
Finally, we observe from Fig.~\ref{fig:snr_rate_power}(b) that the proposed \gls{UT} scheme always outperforms the perfect codeword selection in terms of effective rate.
%%% old version ---
% For the larger codebook, the proposed \gls{UT} scheme still achieves a higher effective rate than the hierarchical search baseline, but the gap is slightly decreased as compared to the linear movement. \begin{change}For the smaller codebook, the average effective rate of the proposed \gls{UT} scheme is slightly lower compared to the perfect hierarchical search baseline.
% This happens since the variance of the \gls{SNR} achieved with the proposed \gls{UT} scheme is comparatively large, due to potentially incorrect codeword selection, leading to large \gls{SNR} degradation.
% Stochastic variations in the \gls{SNR} are detrimental for the effective rate, c.f. \eqref{eq:evaluation:def_gamma}.
% On the other hand, wrong codeword selection is not possible by definition for the perfect hierarchical search (rendering its performance as observed in Fig.~\ref{fig:snr_rate_power}(b) an upper bound for practical schemes based on hierarchical search).\end{change} Finally, we observe from Fig.~\ref{fig:snr_rate_power}(b) that the proposed \gls{UT} scheme always outperforms the perfect codeword selection in terms of effective rate.

\Copy{eval_example_trajectory}{\begin{change}Before evaluating the accuracy of direction extrapolation, in Fig.~\ref{fig:example_angle_preidction}, we show the estimated angles for a short 10 second time window for the \textit{proposed \gls{UT} scheme} and linear movement. The solid lines represent the ground truth direction, the dots represent the estimated direction, and the dashed lines show the predicted direction using a linear prediction polynomial, i.e., $n=1$. One can observe that in the first \gls{TB} from 0 to 1 second, accurate prediction is not possible. This is expected, since at least two direction estimates are required to uniquely fit a first-order polynomial. However, once two observations are available after the second \gls{IDE} block, the linear prediction closely aligns with the user's true movement path.\end{change}}
\Copy{eval_example_trajectory_fig_only}{\begin{figure}
        \vspace{-3mm}
        \centering
        \includegraphics[clip, trim=0cm 0cm 0cm 10mm,width=0.45\textwidth]{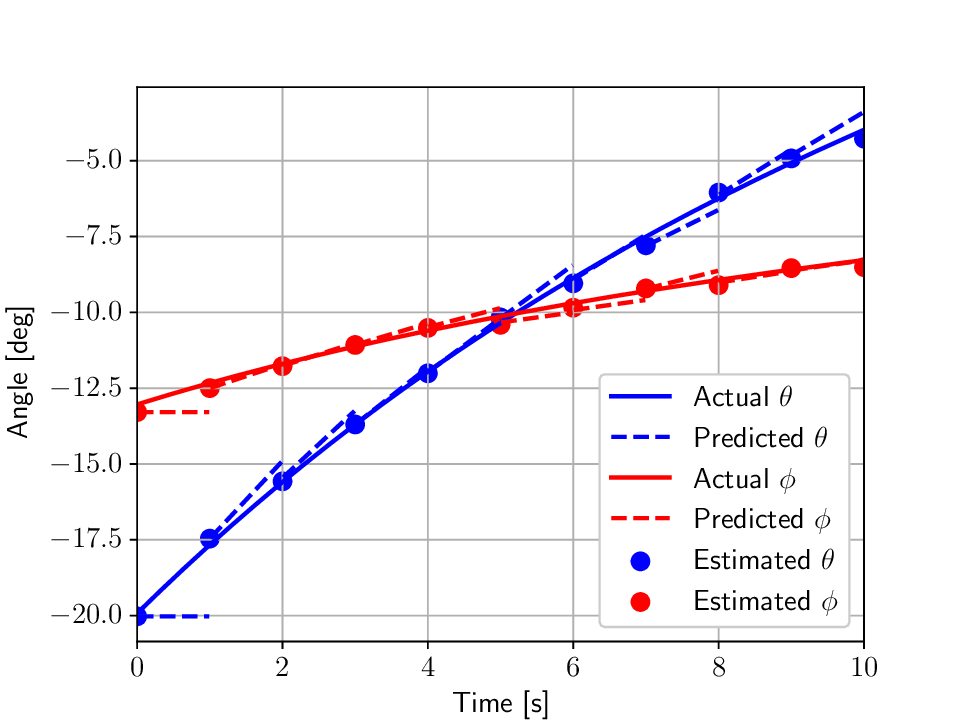}
        \vspace{-2mm}
        \copyablefigurecaptionpluslabel{Example for direction estimations in regular intervals ($T$=1 [s]) and the predicted directions compared to the ground truth values of $\theta$ and $\phi$.}{fig:example_angle_preidction}
        \vspace{-5mm}
\end{figure}}

\Copy{mse_over_time_with_Kalman_fig_only}{
\begin{figure}
    \centering
    \includegraphics[clip, trim=0cm 0cm 0cm 10mm,width=0.45\textwidth]{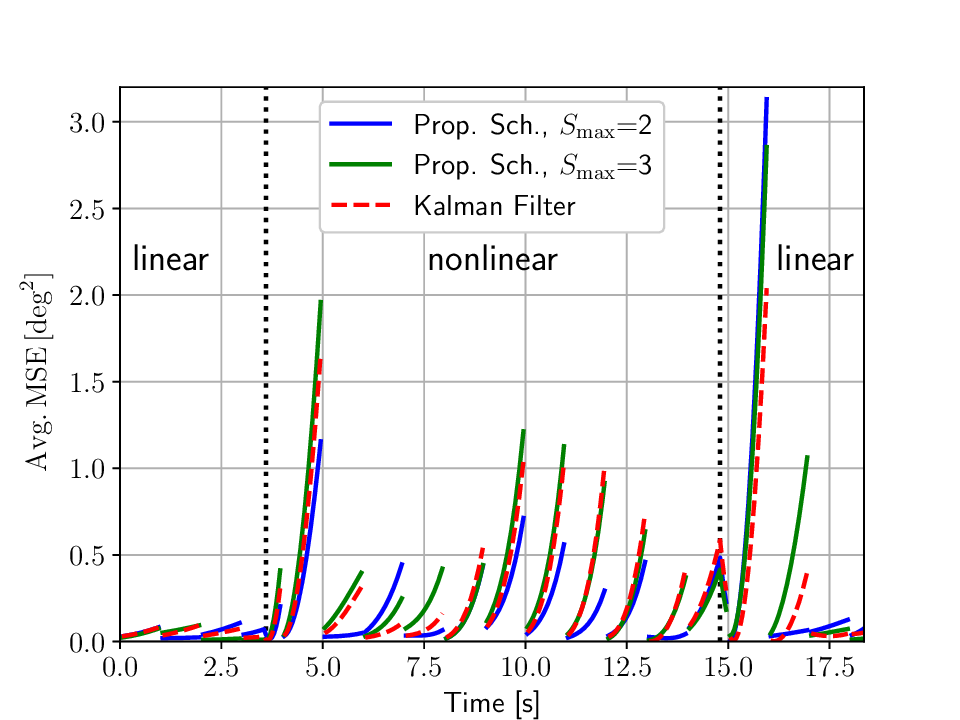}
    \vspace{-2mm}
    \copyablefigurecaptionpluslabel{Direction prediction error over time for an exemplary user trajectory realization.}{fig:mse_over_time}
    \vspace{-5mm}
\end{figure}}
\Copy{mse_over_time_with_Kalman}{Completing the discussion in this section, we now study the specific challenges for trajectory extrapolation in the considered nonlinear movement scenario in more detail.
To this end, Fig.~\ref{fig:mse_over_time} shows the direction prediction error $\left\| \widehat{\boldsymbol\psi}_k(t) - \boldsymbol\psi_{\mathrm{UE,1}}(t) \right\|_2^2$ (averaged over several realizations of the noise) for transmit power $P_\mathrm{TX} = 100 \mathrm{mW}$ for an exemplary nonlinear user trajectory \begin{change} for the proposed trajectory extrapolation scheme and a Kalman filter, which is described in Appendix~\ref{app:kalman_filter}.\end{change} 
We compare three different curves. The proposed scheme is evaluated for $S_\mathrm{max}=2$ (blue) and $S_\mathrm{max}=3$ (green), and, in addition, we consider the Kalman filter baseline (red). For all three schemes, the prediction is regularly updated in the \gls{IDE} sub-blocks, leading to a comparatively low prediction error directly after each \gls{IDE} sub-block, since the obtained direction estimates are very reliable.
Then, the direction prediction error increases during the subsequent extrapolation phase (\gls{CE}, \gls{D}, and \gls{UC} sub-blocks) before decaying again in the next \gls{IDE} sub-block.
During the initial linear movement section, the prediction works seamlessly, resulting in a low error. 
However, directly after transitioning to the nonlinear movement section, previous direction estimates are not informative of the new movement trajectory. This leads to comparatively large errors, that reduce again after sufficiently many new direction estimates are obtained.
Finally, we observe that, after transitioning into the second linear movement section, the prediction error first increases (since, as in the first transition, the trajectory at the point of transition is non-smooth) before reducing again to a low value.
These observations confirm that the transitions between the different movement patterns as well as the nonlinear part of the user trajectory render the prediction less reliable as compared to the linear movement scenario.

Apart from this general performance trend that is identical for all three considered schemes, some differences between the individual schemes can be observed.
For the proposed scheme, the speed at which the trajectory extrapolation adapts to the altered movement patterns (at the transitions from linear to nonlinear movement and vice versa marked by the two dotted lines in Fig.~\ref{fig:mse_over_time}) depends on the number of considered past direction estimates $S_\mathrm{max}$, namely the prediction error reduces quicker after each transition for $S_\mathrm{max}=2$ than for $S_\mathrm{max}=3$. 
\begin{change}The Kalman filter performs similar to the two realizations of the proposed scheme. 
However, as discussed before, the performance of the Kalman filter depends on the proper choice of parameter values (especially the process noise) that may be difficult to select in practice (see Remark \ref{remark_3} in Section \ref{sec:tracking_scheme}) \cite{sharma2011alpha}. Hence, we conclude that the proposed scheme provides a good trade-off between generality and accuracy.\end{change}}% end kalman filter

From the results presented in this section, we conclude that the proposed \gls{UT} scheme still delivers competitive performance both in terms of the average realized \gls{SNR} and the average effective rate even for nonlinear user movement which is notoriously difficult to predict.
Furthermore, it is confirmed that a higher codebook resolution in terms of the number of codewords is required to effectively serve nonlinearly moving users as compared to linear moving ones.

%%%%%%%%%%%%%%%%%%%%
%%%%%%%%%%%%%%%%%%%%
%%%%%%%%%%%%%%%%%%%%
\section{Conclusion}\copyablelabel{sec:conclusion}
In this paper, a novel codebook-based \gls{UT} algorithm has been presented. Specifically, we introduced a peak \gls{ML}-based direction estimation method and a \begin{change}direction\end{change} extrapolation scheme that tracks general user movements by adapting a trajectory estimate to several past direction measurements. The proposed algorithm utilizes the extrapolated trajectory to predict the future directions of the user and anticipates if a codeword change benefits the average achieved \gls{SNR}, improving the average \gls{SNR} compared to baseline schemes with less frequent \gls{IRS} configuration updates. The impact of the employed \gls{IRS} codebook on the direction estimation error has been theoretically analyzed and the results of this analysis were utilized to generate a codebook optimized for direction estimation. \Copy{name_advantages_conclusion}{\begin{change}The proposed user tracking algorithm benefits from the novel codebook design, specifically adapted to the considered direction estimation problem and has a low overhead due to the trajectory extrapolation based updates of the employed codeword between two direction estimates.\end{change}} Our numerical results show that both the proposed \gls{UT} scheme and the proposed optimized codebook design outperform various baseline schemes. Hence, we believe that the proposed schemes present a significant step towards the effective utilization of \gls{IRS}-assisted communication systems for serving mobile users with arbitrary movement trajectories.

\Copy{multi_user_extension}{\begin{change}Interesting topics for further research include the consideration of multiple simultaneously moving users which are served over non-orthogonal channels and the usage of multiple \glspl{IRS} for user tracking and direction estimation.\end{change}} 
\Copy{footnote_alternative_to_polynomial}{\begin{change}Furthermore, we suggest to consider other approaches to replace the polynomial-based trajectory extrapolation scheme to further improve the trajectory extrapolation accuracy. Especially neural network based approaches are interesting, as they may be able to learn the exact street pattern specific to each location.\end{change}}

%%%%%%%%%%%%%%%%%%%%
%%%%%%%%%%%%%%%%%%%%
%%%%%%%%%%%%%%%%%%%%
\appendix
\subsection{Solution of \eqref{eq:codebook_design:omega_opt_v2}}\copyablelabel{app:solution_of_eq:codebook_design:omega_opt_v2}
\Copy{appendix_gradient_descent_phase_shifts}{\begin{change}
The derivative of $\widehat{\mathrm{MSE}}_k$ with respect to each $\rho_i$ is computed as follows
\begin{align}
    \frac{\partial}{\partial \rho_i} & \widehat{\mathrm{MSE}}_k \propto -\int\limits_{\bar{\mathcal{H}}_k} \int\limits_{\bar{\mathcal{H}}_k} || \vec{\psi} - \vec{\psi}' ||_2^2 \nonumber\\
    & \,\times \exp \left(-\frac{||\xi \mathbf{s}||_2^2}{\sigma^2}\sum_{\vec{m} \in \Mk}|\left(\vec{a}^{\mathrm{H}}(\vec{\psi})-\vec{a}^{\mathrm{H}}(\vec{\psi}')\right)\vec{\omega}(\vec{m})|^2 \right)\nonumber\\
    & \,\times \sum_{\vec{m} \in \Mk}\frac{\partial}{\partial \rho_i}|\left(\vec{a}^{\mathrm{H}}(\vec{\psi})-\vec{a}^{\mathrm{H}}(\vec{\psi}')\right)\vec{\omega}(\vec{m})|^2 \,\mathrm{d}\vec{\psi}\,\mathrm{d}\vec{\psi}',\copyablelabel{eq:derivation_mse_k}
\end{align}
where we have omitted the proportionality constant and set without loss of generality $\vec{a}(\vec{\psi}_{\mathrm{t},1}) = \mathbf{1}_{Q^2}$. The derivative in the last term of \eqref{eq:derivation_mse_k} can be further simplified as follows
\begin{align}
    &\frac{\partial}{\partial \rho_i} |\left[\vec{a}(\vec{\psi})-\vec{a}(\vec{\psi}')\right]^{\mathrm{H}}\vec{\omega}(\vec{m})|^2 \nonumber \\
    &= \frac{\partial}{\partial \rho_i} |\left[\vec{a}(\vec{\psi})-\vec{a}(\vec{\psi}')\right]^{\mathrm{H}} \left[\left(\omegaM \odot \vec{\omega}_{m_1}\right) \otimes \left(\omegaM \odot \vec{\omega}_{m_2} \right)\right]|^2 \nonumber \\
    &= \frac{\partial}{\partial \rho_i} |\left[\vec{a}(\vec{\psi})-\vec{a}(\vec{\psi}')\right]^{\mathrm{H}} \left[\left(\omegaM \otimes \omegaM\right) \odot \left(\vec{\omega}_{m_1} \otimes \vec{\omega}_{m_2} \right)\right]|^2.\copyablelabel{eq:codebook_design:simplify_gradient}% \\
\end{align}
Since in \eqref{eq:codebook_design:simplify_gradient} only $\omegaM$ depends on $\rho_i$, $\frac{\partial}{\partial \rho_i} \widehat{\mathrm{MSE}}_k$ can be computed very efficiently using numerical integration.
Then, to find ${\rho}^{*}\left(\omegaM^{(0)}\right)$, gradient descent can be applied as follows
\begin{equation}
    [\rho^{(l+1)}_1,\ldots,\rho^{(l+1)}_Q]^{\mathrm{T}} = [\rho^{(l)}_1,\ldots,\rho^{(l)}_Q]^{\mathrm{T}} - w^{l} \zeta \nabla_{\vec\rho} \widehat{\mathrm{MSE}}_k,\copyablelabel{eq:codebook_design:gradient_descent}
\end{equation}
where $l \in \mathbb{N}_0$, ${\rho}\left(\omegaM^{(0)}\right) = [\rho^{(0)}_1,\ldots,\rho^{(0)}_Q]^{\mathrm{T}}$, $\zeta$ denotes the step size, $w \in (0,1]$, and \eqref{eq:codebook_design:gradient_descent} is applied until the stopping criterion $||w^{l}\zeta \nabla \widehat{\mathrm{MSE}}_k||_2 < \mu$ is achieved for some $\mu > 0$, $l' \in \mathbb{N}_0$. The locally optimal beam $\omegaM^{*}\left(\omegaM^{(0)}\right)$ is then obtained as $\omegaM^{*}\left(\omegaM^{(0)}\right) = \rho^{-1}\left([\rho^{(l')}_1,\ldots,\rho^{(l')}_Q]\right)$.
\end{change}}

%%%%%%%%%%%%%%%%%%%%%%%%%%%%%%%%%%%%%%%%%%%%%%%%%%%%%%%%%%%%%%%%%%%%%%%%%%%%%%%%%%%%%%%%%%%%%%%%%%%%%%%%%%%%%%%%%
\subsection{MUSIC Algorithm}\copyablelabel{app:music_algorithm}
\Copy{appendix_music_algorithm}{\begin{change}
In the following, we describe the \gls{MUSIC} algorithm baseline, which is based on the theory from \cite{wang2023music}. The \gls{MUSIC} algorithm estimates the user's direction based on the pilot signals that are received from codewords $\vec{m}\in\mathcal{M}_k$ as described in Section~\ref{sec:direction_estimation}.
Specifically, in each \gls{IDE} block $k$, all received signals from codewords $\vec{m}\in\mathcal{M}_k$ are collected in vector $\vec{r}_{\vec{m}}$. In \cite{wang2023music}, randomly generated \gls{IRS} configurations are employed, but in Section \ref{sec:performance}, the proposed codebook design and the quadratic codebook design are also evaluated.
Now, for ease of notation of the \gls{MUSIC} algorithm, we denote the vector containing the $n_\mathrm{IDE}$-th pilot symbols of all considered codewords $\vec{m}\in\mathcal{M}_k$ as $\vec{r}_{\mathrm{MUSIC}}^{n_\mathrm{IDE}}$. The sample covariance matrix is then given as follows \cite{wang2023music}
\begin{equation}\copyablelabel{eq:music_S}
    \vspace{-1mm}
    \mathbf{S} = \frac{1}{N_\mathrm{IDE}} \sum_{n_{\mathrm{IDE}}=1}^{N_\mathrm{IDE} } \vec{r}_{\mathrm{MUSIC}}^{n_\mathrm{IDE}} \left(\vec{r}_{\mathrm{MUSIC}}^{n_\mathrm{IDE}}\right)^H .
    \vspace{-1mm}
\end{equation}
The eigenvalue decomposition of $\mathbf{S}$ is given by $\mathbf{S} = \mathbf{U} \boldsymbol{\Lambda} \mathbf{U}^H$, where $\boldsymbol{\Lambda} = [\lambda_1,\lambda_2,...,\lambda_{|\mathcal{M}_{\mathrm{IDE}}|}]$ contains the eigenvalues in decreasing order and $\mathbf{U} = [\vec{u}_1,\vec{u}_2,...,\vec{u}_{|\mathcal{M}_{\mathrm{IDE}}|}]$ contains the eigenvectors. Also, we define $\bar{\mathbf{U}} = [\vec{u}_2,...,\vec{u}_{|\mathcal{M}_{\mathrm{IDE}}|}]$. The \gls{MUSIC} direction estimator is given by \cite{wang2023music}
\begin{equation}\copyablelabel{eq:music_final}
\vspace{-2mm}
    \tilde{\boldsymbol\psi}_k = \arg\max_{\boldsymbol\psi \in \mathcal{H}_k} \frac{1}{\vec{a}_{\mathrm{MUSIC}}^H(\boldsymbol\psi) \bar{\mathbf{U}} \bar{\mathbf{U}}^H \vec{a}_{\mathrm{MUSIC}}(\boldsymbol\psi)},
    \vspace{-0mm}
\end{equation}
where $\vec{a}_{\mathrm{MUSIC}}(\boldsymbol\psi) \in \mathbb{C}^{|\mathcal{M}_k|}$ contains the expected received signals $g_{\vec{m}}(\boldsymbol\psi)\tilde{\xi}(\boldsymbol\psi) [\vec{s}]_{n_{\mathrm{IDE}}}$ of all codewords $\vec{m}\in\mathcal{M}_k$ and we assume that $[\vec{s}]_{n_{\mathrm{IDE}}}$ is identical for all $n_{\mathrm{IDE}}$.
\end{change}}

%%%%%%%%%%%%%%%%%%%%%%%%%%%%%%%%%%%%%%%%%%%%%%%%%%%%%%%%%%%%%%%%%%%%%%%%%%%%%%%%%%%%%%%%%%%%%%%%%%%%%%%%%%%%%%%%%
\subsection{Kalman Filter}\copyablelabel{app:kalman_filter}
\Copy{appendix_kalman_description}{\begin{change}
In this appendix, some details on the Kalman filter algorithm that is used as a baseline in Fig.~\ref{fig:mse_over_time} are provided. The Kalman filter follows the same block structure as described in Section~\ref{sec:system_model:time_block_structure}. The following is based on the Kalman filter theory presented in \cite{kay1993fundamentals}.
In each \gls{TB} $k$, the Kalman filter involves three steps:
\paragraph{Step 1} Measure the direction $\tilde{\boldsymbol\psi}_k$. This is done as described in the proposed approach (see Section \ref{sec:direction_estimation}).
\paragraph{Step 2} Update the current estimate using the state equations. The state update equations are given as follows
\begin{equation}\copyablelabel{eq:kalman_update_K}
\vspace{-1mm}
    \mathbf{K}_k = \mathbf{P}_{k|k-1} \mathbf{H}_K^T \left( \mathbf{H}_k \mathbf{P}_{k|k-1} \mathbf{H}_k^T + \mathbf{R}_k \right)^{-1},
    \vspace{-1mm}
\end{equation}
\begin{equation}\copyablelabel{eq:kalman_update_x}
\vspace{-1mm}
    \vec{x}_k = \vec{x}_{k|k-1} + \mathbf{K}_k \left( \tilde{\boldsymbol\psi}_k - \mathbf{H}_k \vec{x}_{k|k-1} \right),
    \vspace{-1mm}
\end{equation}
and
\begin{equation}\copyablelabel{eq:kalman_update_P}
\vspace{-1mm}
    \mathbf{P}_{k|k} = \left( \mathbf{I}_4 - \mathbf{K}_k \mathbf{H}_k \right) \mathbf{P}_{k|k-1},
    \vspace{-1mm}
\end{equation}
where $\mathbf{K}_k \in \mathbb{R}^{4\times 2}$ is the Kalman gain, $\vec{x}_{k|k-1} = [\theta,\dot{\theta},\phi,\dot{\phi}]^T$ is the state vector containing the direction ($\theta$,$\phi$) and the first derivative of the direction ($\dot{\theta}$,$\dot{\phi}$), the observation matrix is
\begin{equation}\copyablelabel{eq:kalman_H}
\vspace{-1mm}
    \mathbf{H}_k = \mathbf{H} = \begin{bmatrix}
1 & 0 & 0 & 0 \\
0 & 0 & 1 & 0
\end{bmatrix},
\vspace{-1mm}
\end{equation}
and $\mathbf{R}_k\in \mathbb{R}^{2\times 2}$ is the covariance matrix of the measurement noise. Furthermore, $\mathbf{P}_{k|k}\in \mathbb{R}^{4\times 4}$ denotes the covariance matrix of the prediction error of $\vec{x}_k$ after the observation of $\tilde{\boldsymbol\psi}_k$ (a-posteriori), while $\mathbf{P}_{k|k-1}\in \mathbb{R}^{4\times 4}$ is the covariance matrix of the error of $\vec{x}_k$  before the observation of $\tilde{\boldsymbol\psi}_k$ (a-priori).
\paragraph{Step 3} Predict the next state using the equations
\begin{equation}\copyablelabel{eq:kalman_predict_x}
\vspace{-1mm}
    \vec{x}_{k|k-1} = \vec{F}_{k-1} \vec{x}_{k-1},
    \vspace{-1mm}
\end{equation}
and
\begin{equation}\copyablelabel{eq:kalman_predict_P}
\vspace{-1mm}
    \mathbf{P}_{k|k-1} = \mathbf{F}_{k-1} \mathbf{P}_{k-1} \mathbf{F}^T_{k-1} + \mathbf{Q}_{k-1},
    \vspace{-1mm}
\end{equation}
where the state update matrix is
\begin{equation}\copyablelabel{eq:kalman_F}
\vspace{-1mm}
    \mathbf{F}_k = \mathbf{F} = \begin{bmatrix}
1 & T & 0 & 0 \\
0 & 1 & 0 & 0 \\
0 & 0 & 1 & T \\
0 & 0 & 0 & 1 
\end{bmatrix},
\vspace{-1mm}
\end{equation}
and $\mathbf{Q}_{k}$ is the process noise.
It is a known issue that the proper selection of $\mathbf{Q}_k$ is critical for the performance of the Kalman filter \cite{sharma2011alpha}. Theoretically, the process noise should be low in linear movement sections and high in nonlinear movement sections, where the movement trajectory and speed may change, or when the user changes the movement path. 
For the Kalman filter results shown in Fig.~\ref{fig:mse_over_time}, we set $\mathbf{Q}_k = \sigma^2_{\mathrm{K,Q}} \mathbf{I}_4$ and $\mathbf{R}_k = \sigma^2_{\mathrm{K,R}} \mathbf{I}_2$, where we selected $\sigma^2_{\mathrm{K,Q}}$ and $\sigma^2_{\mathrm{K,R}}$ such that the \gls{MSE} is minimized. Note that this optimal selection is not possible in general in practice.
\end{change}}

\renewcommand{\baselinestretch}{0.975}
\bibliographystyle{IEEEtran}
\bibliography{thesis.bib}

\begin{IEEEbiography}[{\includegraphics[width=1in,height=1.25in,clip,keepaspectratio]{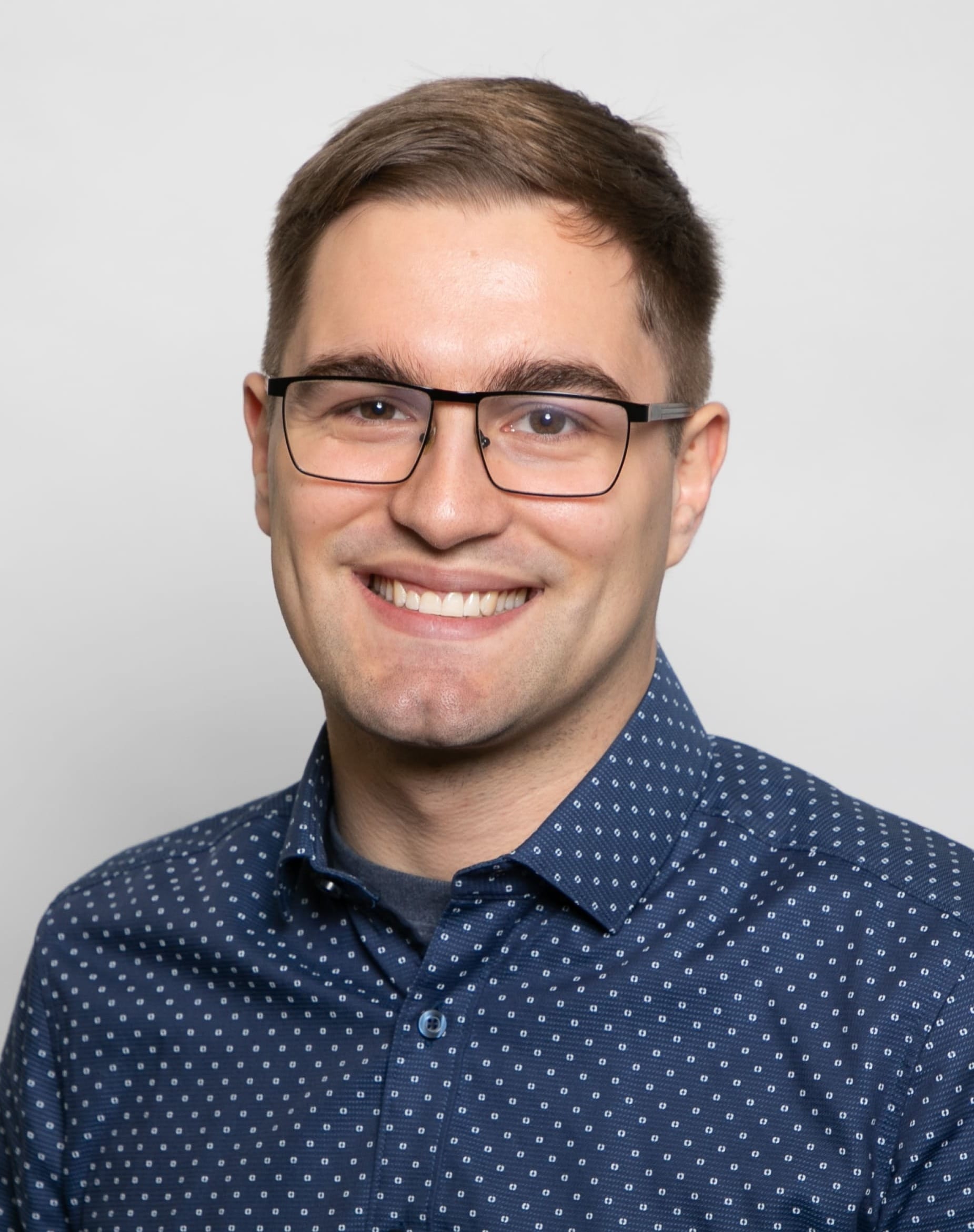}}]
      {Moritz Garkisch} (Graduate Student Member, IEEE) received the {B.Sc.} degree in Electrical Engineering and the {M.Sc.} degree in Advanced Signal Processing and Communications Engineering from the Friedrich–Alexander-Universität Erlangen–Nürnberg, Germany, in 2020 and 2022, respectively, where he is currently pursuing the Ph.D. degree in Electrical Engineering at the Institute for Digital Communication. His main research interests are cellular communication, intelligent reflecting surfaces, sensing/localization, as well as wireless communication in general.
\end{IEEEbiography}

\begin{IEEEbiography}[{\includegraphics[width=1in,height=1.25in,clip,keepaspectratio]{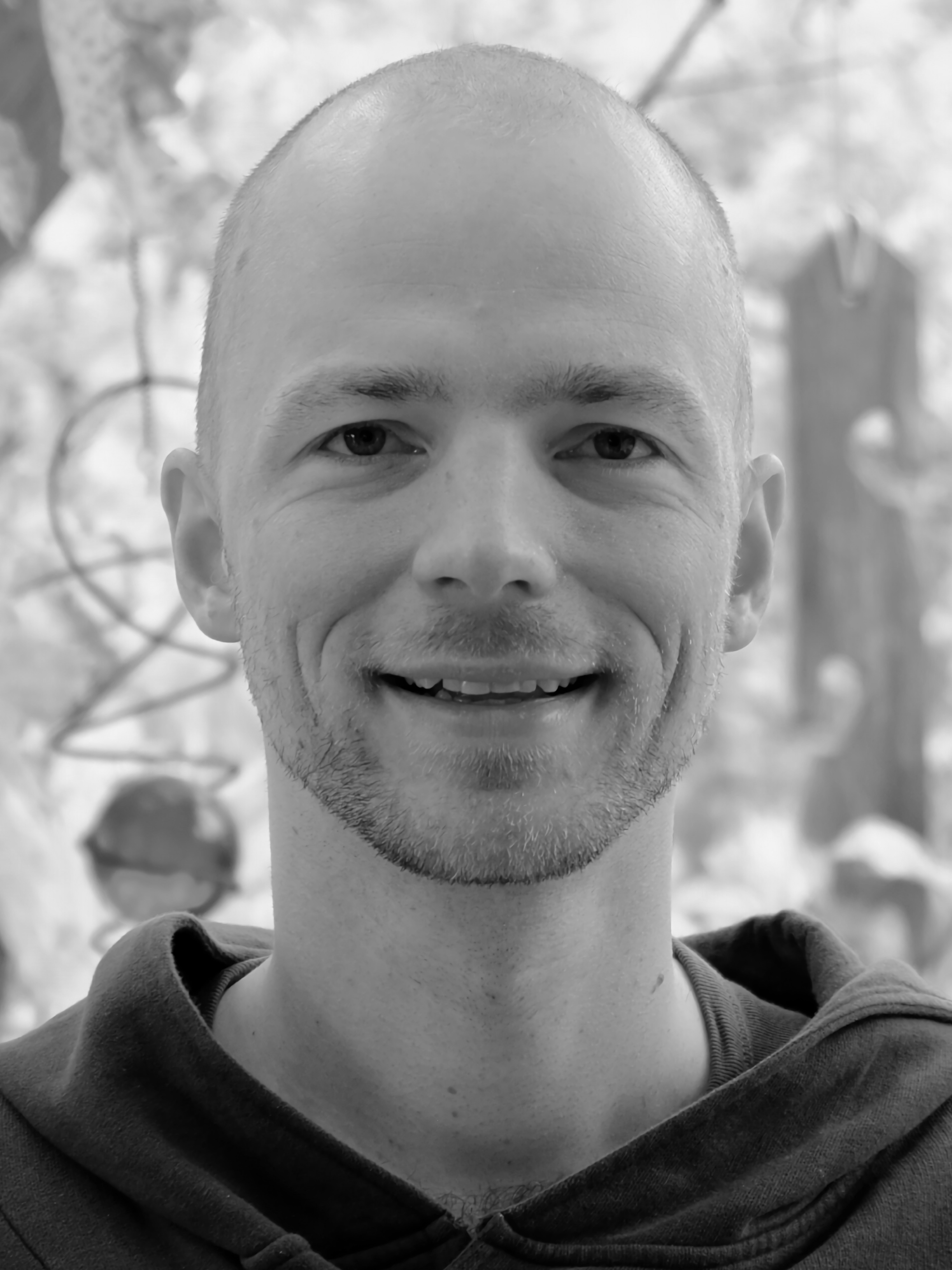}}]
      {Sebastian Lotter} (Member, IEEE) received the M.Sc. and Doctoral degrees from Friedrich–Alexander-Universität Erlangen–Nürnberg in 2019 and 2022, respectively, where he is currently a Postdoctoral Researcher with the Institute for Digital Communication. His research interests include molecular communications, mathematical channel modeling, and communication theory in general. He received the Best PhD Thesis Award from the VDE Bayern in 2023, Best Paper Awards from the IEEE ICC in 2020 and the ACM NanoCOM in 2022, and a Student Travel Grant from the IEEE GLOBECOM in 2022.
\end{IEEEbiography}

\begin{IEEEbiography}[{\includegraphics[width=1in,height=1.25in,clip,keepaspectratio]{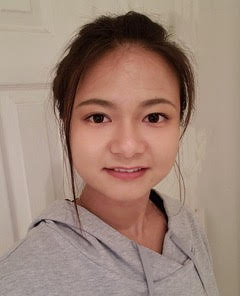}}]
      {Gui Zhou} (Member, IEEE) received the {B.S.} and {M.E.} degrees from the School of Information and Electronics, Beijing Institute of Technology, Beijing, China, in 2015 and 2019, respectively, and the {Ph.D.} degree from the School of Electronic Engineering and Computer Science, Queen Mary University of London, U.K. in 2022. She was a Humboldt Post-Doctoral Research Fellow with the Institute for Digital Communications, Friedrich-Alexander-Universität Erlangen-N\"{u}rnberg (FAU), Erlangen, Germany, from 2022-2024. She is currently a Post-Doctoral Research Fellow in the same university. Her major research interests include channel estimation, transceiver design, integrated sensing and communication, and array signal processing. She is currently an Editor of IEEE Transactions on Communications.
\end{IEEEbiography}

\begin{IEEEbiography}[{\includegraphics[width=1in,height=1.25in,clip,keepaspectratio]{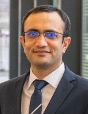}}]
      {Vahid Jamali} (Senior Member, IEEE) received the doctoral degree (with distinctions) from Friedrich-Alexander-Universität Erlangen-Nürnberg (FAU) in 2019. He has been an Assistant Professor with the Technical University of Darmstadt (TUDa), since 2022, leading the Resilient Communication Systems Lab. Prior to joining TUDa, he held academic appointments at Princeton University (2021–2022) and FAU (2019– 2021), as a Post-Doctoral Researcher; and at Stanford University as a Visiting Researcher in 2017. His research interests include wireless and molecular communications. He has served as an Associate Editor of the \textsc{IEEE Transactions on Communications}, \textsc{IEEE Communications Letters}, and \textsc{IEEE Open Journal of the Communications Society} as well as a Vice-Chair for the IEEE ComSoc -- German chapter. He has received several awards for his publications including the Best Paper Awards from the IEEE ICC in 2016, the ACM NanoCOM in 2019, the Asilomar CSSC in 2020, and the IEEE WCNC in 2021; and the Best Journal Paper Award (Literaturpreis) from the German Information Technology Society (ITG) in 2020.
\end{IEEEbiography}
\vfill

\begin{IEEEbiography}[{\includegraphics[width=1in,height=1.25in,clip,keepaspectratio]{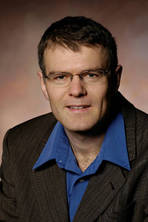}}]
      {Robert Schober} (S’98–M’01–SM’08–F’10) was born in Neuendettelsau, Germany, in 1971. He received the Diplom (Univ.) and Ph.D. degrees in electrical engineering from the Friedrich-Alexander-Universität Erlangen–Nürnberg in 1997 and 2000, respectively. From 2001 to 2002, he was a Post-Doctoral Fellow with the University of Toronto, Canada, sponsored by the German Academic Exchange Service. Since 2002, he has been with The University of British Columbia, Vancouver, Canada, where he is currently a Full Professor. Since 2012, he has been an Alexander von Humboldt Professor and the Chair of Digital Communication with the Friedrich-Alexander-Universität Erlange–Nürnberg, Germany. His research interests include the broad areas of communication theory, wireless communications, and statistical signal processing. Dr. Schober is a fellow of the Canadian Academy of Engineering and the Engineering Institute of Canada. From 2012 to 2015, he served as an Editor-in-Chief of the IEEE TRANSACTIONS ON COMMUNICATIONS. Since 2014, he has been the Chair of the Steering Committee of the IEEE TRANSACTIONS ON MOLECULAR, BIOLOGICAL, AND MULTISCALE COMMUNICATIONS. Furthermore, he is a Member-at-Large of the Board of Governors of the IEEE Communications Society. He has received several awards for his work, including the 2002 Heinz MaierLeibnitz Award of the German Science Foundation, the 2004 Innovations Award of the Vodafone Foundation for Research in Mobile Communications, the 2006 UBC Killam Research Prize, the 2007 Wilhelm Friedrich Bessel Research Award of the Alexander von Humboldt Foundation, the 2008 Charles McDowell Award for Excellence in Research from UBC, a 2011 Alexander von Humboldt Professorship, and a 2012 NSERC E.W.R. Steacie Fellowship. In addition, he has received several best paper awards for his research.
\end{IEEEbiography}

\vfill

\end{document}